\documentclass[12pt]{article}
\usepackage{geometry}             
\usepackage{graphicx}
\usepackage{amssymb}
\usepackage{amsmath}
\usepackage{epstopdf}
\usepackage{subfigure}
\usepackage{cite}
\usepackage{color}
\usepackage{epsf}
\usepackage{appendix}

\usepackage[hyperindex=true,
          pdfstartview=FitH,
          bookmarksnumbered=true,
          bookmarksopen=true,
          citecolor=blue,
          linkcolor=blue,
          colorlinks=true,
          pdfborder=001,
          unicode]{hyperref}

\begin{document}

\title{Holographic complexity of Born-Infeld black holes}
\author{Kun Meng\\
School of Physics and Photoelectric Engineering, Weifang University,\\
Weifang 261061, China\\
School of Physical Science and Technology, Tianjin Polytechnic University, \\
Tianjin 300387, China\\
emails: \href{mailto:mengkun@tjpu.edu.cn}{\it mengkun@tjpu.edu.cn}
 }
\date{}                             
\maketitle

\begin{abstract}
In this paper, according to CA duality, we study complexity growth of Born-Infeld (BI) black holes. As a comparison,  we study action growth of dyonic black holes in Einstein-Maxwell gravity at the beginning. We study action growth of electric BI black holes in dRGT massive gravity, and find BI black holes in massive gravity complexify faster than the Einstein gravity counterparts. We study action growth of the purely electric and  magnetic Einstein-Born-Infeld (EBI) black holes in general dimensions and the dyonic EBI black holes in four-dimensions, and find the manners of action growth are different between electric and  magnetic EBI black holes. In all the gravity systems we considered, we find  action growth rates vanish for the purely magnetic black holes, which is unexpected. In order to ameliorate the situation, we add the boundary term of matter field to the action and discuss the outcomes of the addition.
\end{abstract}

\section{Introduction}
Holographic principle  relates boundary CFT to bulk theory of gravity, through the correspondence one can study the problems of strong coupling CFT on the boundary  through studying weak coupling gravity in the bulk. Remarkable progress has been made in applications of holographic principle in recent years, including applications of holography to study low energy QCD,  hydrodynamics, condensed matter theory\cite{0501128,1101.2451,1103.3022,0803.3295,0810.1563}, etc.

Recently, the combination study of holography and quantum information shed light on understanding of quantum gravity. In the initial work\cite{1306.0533}, Maldacena and Susskind found any pair of entangled black holes are connected by some kind of Einstein-Rosen bridge, i.e., ER=EPR. However, the  ER=EPR duality does not tell how it difficult to transmit information through Einstein-Rosen bridge. Therefore, the concept complexity was introduced. Complexity is the minimal number of simple gates needed to prepare a target state from a reference state. Complexity was originally conjectured to be proportional to the maximum volume of codimension-one surface bounded by the CFT slices, $\mathcal{C}=\frac{V}{Gl}$, which is called CV duality\cite{CV,CV2,1711.10887,1803.06680,1807.06361,1803.08627,1808.08719,1808.10169}. The length scale $l$ is chosen according to situations. In order to eliminate the ambiguities in CV duality, CA duality was proposed\cite{1509.07876,1512.04993}, which states that complexity is proportional to the action in Wheeler-DeWitt(WDW) patch, $\mathcal{C}=\frac{I}{\pi\hbar}$. CA duality does not involve any ambiguities encountered in CV duality but preserves all the nice features of CV duality. CA duality have passed the tests of shock wave and tensor network.

According to the definition, complexity growth rate is the speed of quantum computations. Considering black holes are  the densest memories\cite{9310026,9409089,Bekenstein}, it is conjectured that black holes are the fastest computers in nature. There exist a bound for the speed of quantum computation. Inspired by Margolus-Levitin bound\cite{9710043}
\begin{align}
\textmd{orthogonality}\; \textmd{time}\geq \frac{\pi \hbar}{2\langle E\rangle},\nonumber
\end{align}
which gives the minimal time needed for a state evolving to an orthogonal state, Lloyd proposed a bound on the speed of computation\cite{Lloyd}. Brown and collaborators generalized Lloyd's bound and conjectured that there exists a bound on the growth rate of complexity, which is
\begin{align}
\frac{d\mathcal{C}}{dt}\leq \frac{2E}{\pi\hbar}.\label{complexitybound}
\end{align}
Calculations show that static neutral black holes saturate the bound.

It is natural to study complexity growth of black holes in different gravity systems to examine CA duality and  Lloyd's bound.
For the progresses in this subject please refer to\cite{1512.04993,1606.08307,1702.06766,1703.10006,1703.06297,1712.09826,1703.10468,1612.03627,1801.03638,1806.06216,1702.06796,1806.10312,1706.03788,1803.02795,1710.05686,1808.09917,1610.05090,1804.07410,1805.07262,
1810.00758,1708.01779,1808.00067}.
In this paper, we intend to study complexity growth of BI black holes, since we are interested in the effects of nonlinearity of BI theory on complexity growth. The inner horizon of a BI black hole may turn into a curvature singularity due to perturbatively unstability\cite{1702.06766}, which implies a BI black hole may possess a single horizon\cite{1311.7299,1712.08798,1804.10951}. It's interesting to study the differences in complexity between BI black holes with single horizon and  AdS-Schwarzschild  black holes, although the casual structures of them are identical. Since the magnetic black holes have been studied rarely, in this paper we will pay much attention to the magnetic black holes, and make a comparison of the effects between electric and magnetic charges on action growth. As we will see in the following, action growth of magnetic BI black holes exhibit some specific properties that are not found in the electric ones. We  are also interested in studying action growth of BI black holes in massive gravity and study the effects of graviton mass.

Recently, the authors of refs.\cite{1901.00014,1905.06409,1905.07576} found that, action growth rates vanish for purely magnetic black holes in four dimensions. Which is unexpected since the expected late-time result $\frac{dI}{dt}\sim TS$ and electric-magnetic duality cannot be restored. Similar result was also found in this paper. In order to ameliorate the situation, a boundary term of matter field was proposed to be included to the action. In this paper, we add the boundary term of matter field proper to the gravity systems we considered and discuss the outcomes of the addition of the boundary term.

The paper is organized as, we study action growth of dyonic black holes in Einstein-Maxwell gravity in section \ref{section2}, BI black holes in massive gravity in section \ref{section3}, and EBI black holes in section \ref{section4}. In all the gravity systems we considered, we add the boundary term of matter field and discuss the outcomes of the addition. We summarize our calculations in the last section.


\section{Dyonic black holes of Einstein-Maxwell gravity\label{section2}}
In order to make a comparison between action growth of BI black holes and that of black holes in Einstein-Maxwell gravity,  we first study dyonic black holes in Einstein-Maxwell gravity in this section.
The action of the theory reads
\begin{align}
I=\frac{1}{16\pi}\int d^dx\sqrt{-g}\left[R-2\Lambda-\frac{1}{4}F_{\mu\nu}F^{\mu\nu}\right].\label{EinsteinMax}
\end{align}
After taking variations of the metric and electromagnetic field, the field equations are given by
\begin{align}
G_{\mu\nu}+\Lambda g_{\mu\nu}=\frac{1}{2}F_{\mu\lambda}&F_{\nu}^{\;\;\lambda}-\frac{1}{8}F_{\alpha\beta}F^{\alpha\beta}g_{\mu\nu},\label{eommaxgrav}\\
\nabla_\mu F^{\mu\nu}&=0.\label{eommax}
\end{align}
We take the metric and field strength ansatz for AdS planar black holes in $d=2n+2$ dimensions as
\begin{align}
ds^2&=-f(r)dt^2+\frac{dr^2}{f(r)}+r^2\left(dx_1^2+dx_2^2+\cdots+dx_{2n-1}^2+dx_{2n}^2\right),\nonumber\\
F&=\Phi'(r)dr\wedge dt+p(dx_1\wedge dx_2+\cdots+dx_{2n-1}\wedge dx_{2n}).\label{strengthansatz}
\end{align}
Solving equations of motion (e.o.m.) of the Maxwell field (\ref{eommax}) we obtain
\begin{align}
\Phi(r)=\int_r^\infty \frac{q dr}{r^{2n}}.
\end{align}
Now we are able to solve the e.o.m. of metric (\ref{eommaxgrav}), and obtain the dyonic black hole solution
\begin{align}
f(r)=-\frac{\mu}{r^{d-3}}-\frac{2\Lambda}{(d-1)(d-2)}r^2+\frac{q^2}{2(d-2)(d-3)r^{2(d-3)}}-\frac{p^2}{4(d-5)r^2}.\label{dyonicBHmax}
\end{align}

\begin{figure}[h]
\begin{center}
\includegraphics[width=.40\textwidth]{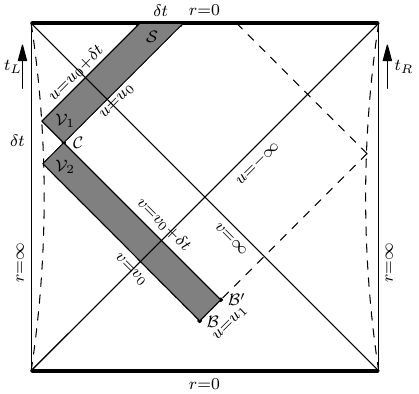}
\includegraphics[width=.40\textwidth]{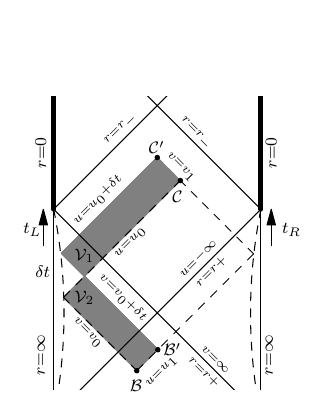}
\end{center}
\caption{Wheeler-DeWitt (WDW) patch of  black holes. The left panel represents WDW patch of black holes with single  horizon, the right panel represents WDW patch of black holes with double horizons.}
\label{fig1}
\end{figure}

According to CA duality, complexity is proportional to action in WDW patch. Since WDW patch is in general non-smooth, as shown by Fig.\ref{fig1}, we employ the method proposed in \cite{1609.00207,9403018} to calculate the action, which is given by
\begin{align}
I_{tot}=\int_{\mathcal{V}}(R-2\Lambda+\mathcal{L}_{mat})\sqrt{-g}dV+2\Sigma_{T_i}\int_{\partial\mathcal{V}_{T_i}}Kd\Sigma+2\Sigma_{S_i}sign(S_i)\int_{\partial\mathcal{V}_{S_i}}
Kd\Sigma\nonumber\\
+2\Sigma_{N_i}sign(N_i)\int_{\partial\mathcal{V}_{N_i}}\kappa dSd\lambda+2\Sigma_{j_i}sign(j_i)\oint \eta_{j_i}dS+2\Sigma_{m_i}sign(m_i)\oint a_{m_i}dS.\label{Itot}
\end{align}
Where $S_i$, $T_i$ and $N_i$ labels spacelike, timelike and null boundary respectively. $K$ is the Gibbons-Hawking term. $\kappa$ measures the failure of $\lambda$ to be an affine parameter on the null generators. $\eta_{j_i}$ is the joint term between non-null hypersurfaces. $a_{m_i}$ is the joint term between null and other types of surfaces. The signatures $sign(N_i), sign(j_i), sign(m_i)$ are determined through the requirement that the gravitational action is additive.

To proceed the calculations, it's convenient to introduce  null coordinates
\begin{align}
du\equiv dt+f^{-1}dr,\;\;\;\;\;\;dv\equiv dt-f^{-1}dr.\label{nullcoord}
\end{align}
Under the null coordinates the metric becomes
\begin{align}
ds^2&=-fdu^2+2dudr+r^2h_{ij}dx^idx^j,
\end{align}
or
\begin{align}
ds^2&=-fdv^2-2dvdr+r^2h_{ij}dx^idx^j.
\end{align}

Since  the dyonic black holes (\ref{dyonicBHmax})  possess both inner and outer horizons, we only need the right panel of Fig.\ref{fig1} to calculate action growth.
From the right panel of Fig.\ref{fig1} we obtain the bulk contribution to $\delta I$
\begin{align}
I_{\mathcal{V}_1}&=\frac{1}{16\pi}\omega_2^n\int_u^{u+\delta t}du\int_{r_{-}}^{r_{max}} dr r^{d-2}\mathcal{L}_{bulk},\nonumber\\
I_{\mathcal{V}_2}&=\frac{1}{16\pi}\omega_2^n\int_v^{v+\delta t}dv\int_{r_{+}}^{r_{max}} dr r^{d-2}\mathcal{L}_{bulk},
\end{align}
where $r_{max}$ is the UV cutoff, and $\omega_2\equiv\int dx_1dx_2=\cdots=\int dx_{2n-1}dx_{2n}$. With the e.o.m. (\ref{eommaxgrav}), we have
\begin{align}
\mathcal{L}_{bulk}
=\frac{4\Lambda}{d-2}+\frac{q^2}{(d-2)r^{2(d-2)}}-\frac{p^2}{2r^4}.
\end{align}
Therefore, the bulk contribution to total action growth is
\begin{align}
I_{\mathcal{V}_1}-I_{\mathcal{V}_2}&=\frac{1}{16\pi}\omega_2^n\delta t\bigg[\frac{4\Lambda r_{+}^{d-1}}{(d-1)(d-2)}-\frac{p^2r_{+}^{d-5}}{2(d-5)}
-\frac{q^2}{(d-2)(d-3)r_{+}^{d-3}}\bigg]\bigg|_{r_{-}}^{r_{+}}.
\end{align}
As shown by the right panel of Fig.\ref{fig1}, there are four joints between null surfaces contributing to $\delta I$. Actions of the joints are given by
\begin{align}
I_{\mathcal{B}'\mathcal{B}}&=\frac{1}{16\pi}\left[2\oint_{\mathcal{B}'}adS-2\oint_{\mathcal{B}}adS\right]\nonumber\\
&=\frac{1}{8\pi}\omega_2^n\left[h(r_{\mathcal{B}'})-h(r_{\mathcal{B}})\right],
\end{align}
where $a=\ln\left(-\frac{1}{2}k\cdot\bar{k}\right)$, with $k$ being the null normal to the hypersurface  $v=const$ and  $\bar{k}$ being the null normal to the hypersurface $u=const$. For the affinely parametrized expressions $k_\alpha=-c\partial_\alpha v$ and $\bar{k}_\alpha=\bar{c}\partial_\alpha u$, we have $a=-\ln\left(-f/(c\bar{c})\right)$, therefore $h(r)=-r^{d-2}\ln\left(-f/(c\bar{c})\right)$. Using $dr=-\frac{1}{2}f\delta t$, we obtain
\begin{align}
I_{\mathcal{B}'\mathcal{B}}=\frac{1}{16\pi}\omega_2^n\delta t\left[r^{d-2}f'+(d-2)r^{d-3}f\ln\left(\frac{-f}{c\bar{c}}\right)\right]\bigg|_{r=r_{\mathcal{B}}}.
\end{align}
At late times, $r_{\mathcal{B}}\rightarrow r_{+}$, $r_{\mathcal{C}}\rightarrow r_{-}$, we have
\begin{align}
I_{\mathcal{B}'\mathcal{B}}=\frac{1}{16\pi}\omega_2^nr^{d-2}f'(r)\big|_{r_{+}}\delta t,\;\;\;\;I_{\mathcal{C}'\mathcal{C}}=-\frac{1}{16\pi}\omega_2^nr^{d-2}f'(r)\big|_{r_{-}}\delta t.\label{4joints}
\end{align}
The two terms in (\ref{4joints}) together give rise to
\begin{align}
I_{\mathcal{B}'\mathcal{B}}+I_{\mathcal{C}'\mathcal{C}}=&\frac{1}{16\pi}\omega_2^n\delta t\bigg[-\frac{4\Lambda r_{+}^{d-1}}{(d-1)(d-2)}+\frac{p^2r_{+}^{d-5}}{2(d-5)}
-\frac{q^2}{(d-2)r_{+}^{d-3}}\bigg]\bigg|_{r_{-}}^{r_{+}}.
\end{align}

The total action growth is obtained by sum of all the contributions
\begin{align}
\frac{\delta I}{\delta t}&=\frac{1}{16\pi}\left[-\frac{q^2}{(d-3)r_{+}^{d-3}}+\frac{q^2}{(d-3)r_{-}^{d-3}}\right]\nonumber\\
&=\left[(M-Q_e\Phi_e)_{+}-(M-Q_e\Phi_e)_{-}\right].\label{IgdyonicMax}
\end{align}
Where $Q_e=\frac{q}{16\pi}, \Phi_e=\frac{q}{(d-3)r_{\pm}^{d-3}}$ are electric charge and potential respectively.  It can be seen that, similar to the four-dimensional case\cite{1901.00014}, magnetic charge does not contribute to action growth, which implies action growth rates vanish for purely magnetically charged black holes. In order to restore the late-time result $\frac{dI}{dt}\sim TS$ for purely magnetic black holes and electric-magnetic duality in four dimensions, the action of Einstein-Maxwell theory may be modified by a boundary term of Maxwell field.

The Maxwell boundary term which is considered to be included to the action reads\cite{1901.00014}
\begin{align}
I_{\mu Q}&=\frac{\gamma}{16\pi}\int_{\partial\mathcal{M}} d\Sigma_\mu F^{\mu\nu}A_\nu.\label{boundaryMax}
\end{align}
This term does not affect the field equations but only alter the boundary conditions in the variational principle of Maxwell field. A well-posed variational principle requires Dirichlet boundary condition $\delta A_a=0$ for the original Maxwell action in (\ref{EinsteinMax}), while after adding the boundary term (\ref{boundaryMax}), it requires Neumann boundary condition $n^\mu\partial_\mu\delta A_a=0$ for $\gamma=1$ and mixed boundary conditions for  general $\gamma$. Addition of the boundary term (\ref{boundaryMax})  is natural when studying thermodynamics or Euclidean action, it produces the Legendre transformation from a grand canonical ensemble with fixed chemical potential to a canonical one with fixed charge. Using the field equations $\nabla_\mu F^{\mu\nu}=0$ and Stokes' theorem the boundary term (\ref{boundaryMax}) can be rewritten as
\begin{align}
I_{\mu Q}=\frac{\gamma}{32\pi}\int_{\mathcal{M}} d^dx\sqrt{-g}F^{\mu\nu}F_{\mu\nu}.
\end{align}
Action growth for the dyonic black holes now becomes
\begin{align}
\frac{\delta I}{\delta t}&=\left[M-(1-\gamma)Q_e\Phi_e-\gamma Q_m\Phi_m\right]_{+}-\left[M-(1-\gamma)Q_e\Phi_e-\gamma Q_m\Phi_m\right]_{-}.\label{dyonMaxadd}
\end{align}
This result takes the identical form with that of the four-dimensional black hole\cite{1901.00014}. One sees that, if we take $\gamma=\frac{1}{2}$, then electric and magnetic charges contribute to action growth on equal footing, this agrees with electric-magnetic duality in four dimensions. If we take $\gamma=1$, contrary to the $\gamma=0$ case, only magnetic charge contributes to action growth, i.e., action growth rates vanish at late times for purely electrically charged black holes in this case.

\section{BI black holes in massive gravity\label{section3}}
In this section, we study action growth of purely electrically charged BI black holes in massive gravity and discuss the effects of graviton mass.
The action of Einstein massive gravity is given by\cite{1508.01311}
\begin{align}
I=\frac{1}{16\pi}\int d^dx\sqrt{-g}\left[R-2\Lambda+\mathcal{L}(\mathcal{F})+m^2\sum_{i=1}^4c_i\mathcal{U}_i(g,f)\right],
\end{align}
where $f$ is a fixed symmetric rank-2 tensor. $c_i$ are constants and $\mathcal{U}_i$ are symmetric polynomials of the eigenvalues of matrix $\mathcal{K}^\mu_{\;\nu}\equiv\sqrt{g^{\mu\alpha}f_{\alpha\nu}}$
\begin{align}
&\mathcal{U}_1=[\mathcal{K}],\;\;\;\mathcal{U}_2=[\mathcal{K}]^2-[\mathcal{K}^2],\;\;\;\mathcal{U}_3=[\mathcal{K}]^3-3[\mathcal{K}][\mathcal{K}^2]+2[\mathcal{K}^3],\nonumber\\
&\mathcal{U}_4=[\mathcal{K}]^4-6[\mathcal{K}^2][\mathcal{K}]^2+8[\mathcal{K}^3][\mathcal{K}]+3[\mathcal{K}^2]^2-6[\mathcal{K}^4].
\end{align}
$\mathcal{L}(\mathcal{F})$ is the Lagrangian density of BI  theory
\begin{align}
\mathcal{L}(\mathcal{F})=4\beta^2\left(1-\sqrt{1+\frac{F^{\rho\sigma}F_{\rho\sigma}}{2\beta^2}}\right).\label{LFm}
\end{align}
Taking variations of the metric and electromagnetic field one obtains the field equations
\begin{align}
G_{\mu\nu}+\Lambda g_{\mu\nu}-&\frac{1}{2}g_{\mu\nu}\mathcal{L}(\mathcal{F})-\frac{2F_{\mu\lambda}F_\nu^{\;\lambda}}{\sqrt{1+\frac{F^{\rho\sigma}F_{\rho\sigma}}{2\beta^2}}}+m^2\chi_{\mu\nu}=0,\label{eommg}\\
&\partial_\mu\left(\frac{\sqrt{-g}F^{\mu\nu}}{\sqrt{1+\frac{F^{\rho\sigma}F_{\rho\sigma}}{2\beta^2}}}\right)=0,\label{eomma}
\end{align}
where
\begin{align}
\chi_{\mu\nu}=&-\frac{c_1}{2}\left(\mathcal{U}_1g_{\mu\nu}-\mathcal{K}_{\mu\nu}\right)-\frac{c_2}{2}\left(\mathcal{U}_2g_{\mu\nu}-2\mathcal{U}_1\mathcal{K}_{\mu\nu}
+2\mathcal{K}^2_{\mu\nu}\right)-\frac{c_3}{2}\big(\mathcal{U}_3g_{\mu\nu}-3\mathcal{U}_2\mathcal{K}_{\mu\nu}\nonumber\\
&6\mathcal{U}_1\mathcal{K}^2_{\mu\nu}-6\mathcal{K}^3_{\mu\nu}\big)-\frac{c_4}{2}\left(\mathcal{U}_4g_{\mu\nu}-4\mathcal{U}_3\mathcal{K}_{\mu\nu}+12\mathcal{U}_2\mathcal{K}^2_{\mu\nu}
-24\mathcal{U}_1\mathcal{K}^3_{\mu\nu}+24\mathcal{K}^4_{\mu\nu}\right).
\end{align}

We take the static metric ansatz
\begin{align}
ds^2=-f(r)dt^2+\frac{dr^2}{f(r)}+r^2h_{ij}dx^idx^j,
\end{align}
where $h_{ij}dx^idx^j$ is the line element of codimension-two hypersurface with constant curvature.
Using the reference metric
\begin{align}
f_{\mu\nu}=diag(0,0,h_{ij}),
\end{align}
the $\mathcal{U}_i$'s can be expressed as
\begin{align}
\mathcal{U}_1=\frac{d_2}{r},\;\;\;\;\;\;
\mathcal{U}_2=\frac{d_2d_3}{r^2},\;\;\;\;\;\;
\mathcal{U}_3=\frac{d_2d_3d_4}{r^3}, \;\;\;\;\;\;
\mathcal{U}_4=\frac{d_2d_3d_4d_5}{r^4},
\end{align}
where $d_i\equiv d-i$ is introduced for convenience. Under the assumption of electrostatic potential $A_\mu=\Phi(r)\delta_\mu^0$,
we solve the e.o.m of $A_\mu$ and find the only non-vanishing component of strength tensor is
\begin{align}
F_{tr}=\frac{\sqrt{d_2d_3}q}{r^{d_2}\sqrt{1+\frac{d_2d_3q^2}{\beta^2r^{2d_2}}}}.
\end{align}
Substituting the above results into (\ref{eommg}), one obtains the black hole solution
\begin{align}
f(r)=&k-\frac{m_0}{r^{d_3}}+\frac{4\beta^2-2\Lambda}{d_1d_2}r^2-\frac{4\beta^2r^2}{d_1d_2}\sqrt{1+\Gamma}+\frac{4d_2q^2}{d_1r^{2d_3}}\mathcal{H}\nonumber\\
&+m^2\left(\frac{c_1r}{d_2}+c_2+\frac{d_3c_3}{r}+\frac{d_3d_4c_4}{r^2}\right),\label{massiveBH}
\end{align}
with
\begin{align}
\Gamma=\frac{d_2d_3q^2}{\beta^2r^{2d_2}},\;\;\;\;\;\;\mathcal{H}=\sideset{_2}{_1}{\mathop{F}}\left[\frac{1}{2},\frac{d_3}{2d_2},\frac{3d-7}{2d_2},-\Gamma\right].\label{gammaH}
\end{align}

The details of geometry and thermodynamics of the black hole (\ref{massiveBH}) can be found in \cite{1508.01311}. The black hole may possess single or double horizons.
Let's first calculate action growth of the single-horizoned black hole. The left panel of Fig.\ref{fig1} shows us the Penrose diagram of this type of black holes. From the panel it is easy to see that, the $\eta$ terms in (\ref{Itot}) vanish because there are no joints between non-null hypersurfaces in WDW patch. It is natural to require all null segments to be affine parametrized, therefore all $\kappa$ terms in (\ref{Itot}) vanish too. Considering time transition symmetry, the left contributions to $\delta I=I(t_0+\delta t)-I(t_0)$ are
\begin{align}
\delta I=I_{\mathcal{V}_1}-I_{\mathcal{V}_2}-2\int_{\mathcal{S}}Kd\Sigma+2\oint_{\mathcal{B}'}adS-2\oint_{\mathcal{B}}adS
\end{align}

With the e.o.m (\ref{eommg}), we obtain the expression of bulk Lagrangian
\begin{align}
\mathcal{L}_{bulk}=\frac{4\Lambda}{d_2}-\frac{8\beta^2}{d_2}\left(1-\sqrt{1+\frac{d_2d_3q^2}{\beta^2r^{2d_2}}}\right)-\frac{m^2}{d_2}
\left[\frac{d_2c_1}{r}-\frac{d_2d_3d_4c_3}{r^3}-2\frac{d_2d_3d_4d_5c_4}{r^4}\right].
\end{align}
Thus  action of the bulk region $\mathcal{V}_1$  is given by
\begin{align}
I_{\mathcal{V}_1}&=\frac{1}{16\pi}\Omega_{d_2}\int_u^{u+\delta t}du\int_\epsilon^{\rho(u)} dr r^{d_2}\mathcal{L}_{bulk},\nonumber\\
&=\frac{1}{16\pi}\Omega_{d_2}\delta t\left[\frac{4\Lambda}{d_1d_2}r^{d_1}-2F(r)-m^2\left(\frac{c_1}{d_2}r^{d_2}-c_3d_3r^{d_4}-2c_4d_3d_4r^{d_5}\right)\right]\bigg|_\epsilon^{\rho(u)}.
\end{align}
where $\Omega_{d_2}$ is the volume of the codimension-two hypersurface, $r=\rho(u)$ represents the $v=v_0+\delta t$ surface, and $F(r)\equiv \int dr r^{d_2}\frac{4\beta^2}{d_2}\left(1-\sqrt{1+\frac{d_2d_3q^2}{\beta^2r^{2d_2}}}\right)$ is introduced for convenience.
Similarly, action of the  bulk region $\mathcal{V}_2$ is given by
\begin{align}
I_{\mathcal{V}_2}&=\frac{1}{16\pi}\Omega_{d_2}\int_v^{v+\delta t}dv\int_{\rho_1(v)}^{\rho_0(v)} dr r^{d_2}\mathcal{L}_{bulk},\nonumber\\
&=\frac{1}{16\pi}\Omega_{d_2}\delta t\left[\frac{4\Lambda}{d_1d_2}r^{d_1}-2F(r)-m^2\left(\frac{c_1}{d_2}r^{d_2}-c_3d_3r^{d_4}-2c_4d_3d_4r^{d_5}\right)\right]\bigg|_{\rho_1(v)}^{\rho_0(v)},
\end{align}
where $r=\rho_{0(1)}(v)$ represents the $u=u_{0(1)}$ surface. In the late-time limit, $\rho_{1}(v)\rightarrow r_{+}$, we have
\begin{align}
I_{\mathcal{V}_1}-I_{\mathcal{V}_2}=\frac{1}{16\pi}\Omega_{d_2}\delta t\left[\frac{4\Lambda}{d_1d_2}r_{+}^{d_1}-2F(r_{+})+2F(0)-m^2\left(\frac{c_1}{d_2}r_{+}^{d_2}-c_3d_3r_{+}^{d_4}-2c_4d_3d_4r_{+}^{d_5}\right)\right],\label{bulkm}
\end{align}
where we have taken $\epsilon\rightarrow0$. Action of the spacelike surface $\mathcal{S}$  is
\begin{align}
I_{\mathcal{S}}&=-\frac{1}{8\pi}\int Kd\Sigma\nonumber\\
&=\frac{1}{16\pi}\Omega_{d_2}\delta td_1\left[m_0-F(0)\right].\label{Km}
\end{align}
where we have used the expression $K=\frac{-1}{r^{d_2}}\frac{d}{dr}\left(r^{d_2}\sqrt{-f}\right)$ for extrinsic curvature and taken the $r\rightarrow0$ limit. At late times, $r_{\mathcal{B}}\rightarrow r_{+}$, the joint terms are given by
\begin{align}
I_{\mathcal{B}'\mathcal{B}}
=&\frac{1}{16\pi}\Omega_{d_2}\delta t\bigg[d_3m_0-\frac{4\Lambda}{d_1d_2}r_{+}^{d_1}-d_3F(r_{+})+r_{+}F'(r_{+})\nonumber\\
&+m^2\left(\frac{c_1}{d_2}r_{+}^{d_2}-c_3d_3r_{+}^{d_4}-2c_4d_3d_4r_{+}^{d_5}\right)\bigg].\label{jointm}
\end{align}

Collecting our calculations, adding up all the contributions  to $\delta   I$ (\ref{bulkm}), (\ref{Km}) and (\ref{jointm}), we finally arrive at
\begin{align}
\frac{\delta I}{\delta t}=2M-Q\Phi-C,\label{onemassive}
\end{align}
where $M, Q, \Phi$ are respectively mass, electric charge and potential of the black hole
\begin{align}
&M=\frac{d_2 m_0\Omega_{d-2}}{16\pi},\;\;\;\;\;\;\;\;\;\;
Q=\frac{\sqrt{d_2d_3}q}{4\pi},\nonumber\\
\Phi&=\sqrt{\frac{d_2}{d_3}}\frac{q}{r_{+}^{d_3}}\sideset{_2}{_1}{\mathop{F}}\left[\frac{1}{2},\frac{d_3}{2d_2},\frac{3d-7}{2d_2},-\frac{d_2d_3q^2}{\beta^2r_{+}^{2d_2}}\right],
\end{align}
and
\begin{align}
C=\frac{\beta^2d_3}{8\pi d_1d_2}\left(d_2d_3\right)^{\frac{d_1}{2d_2}}\left(\frac{q}{\beta}\right)^{\frac{d_1}{d_2}}\frac{\Gamma\left(1/(2d_2)\right)\Gamma\left(d_3/(2d_2)\right)}{\Gamma(1/2)}
\end{align}

\begin{figure}[h]
\begin{center}
\includegraphics[width=.55\textwidth]{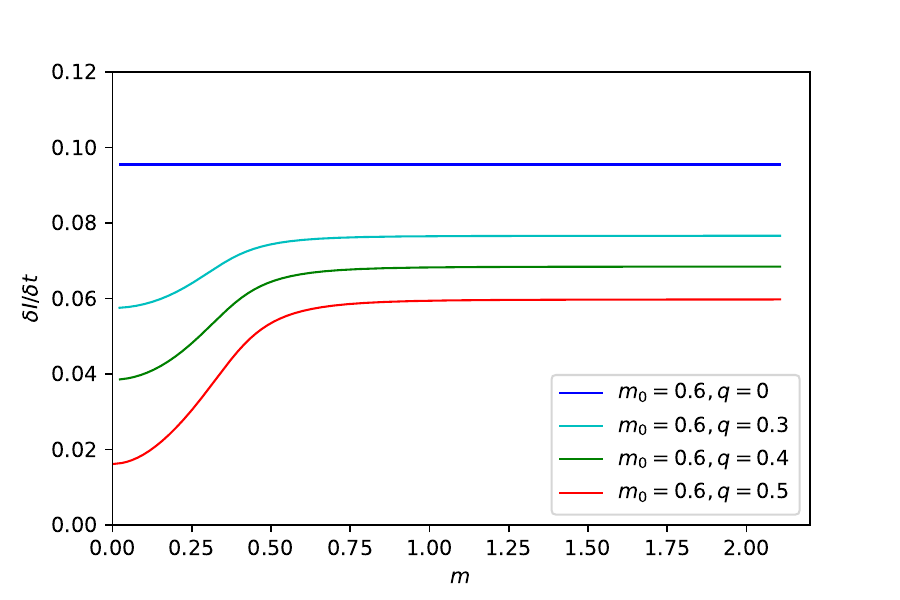}
\includegraphics[width=.45\textwidth]{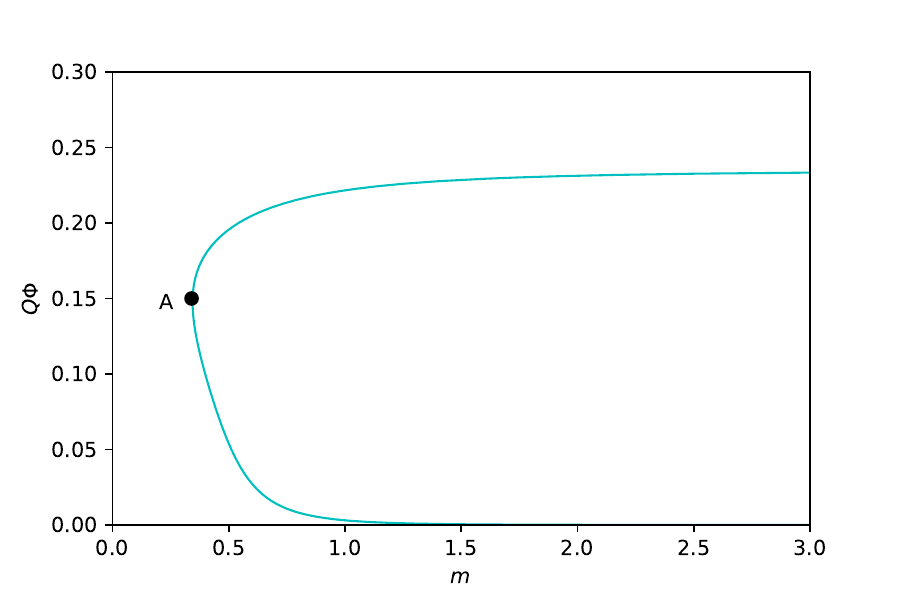}
\includegraphics[width=.45\textwidth]{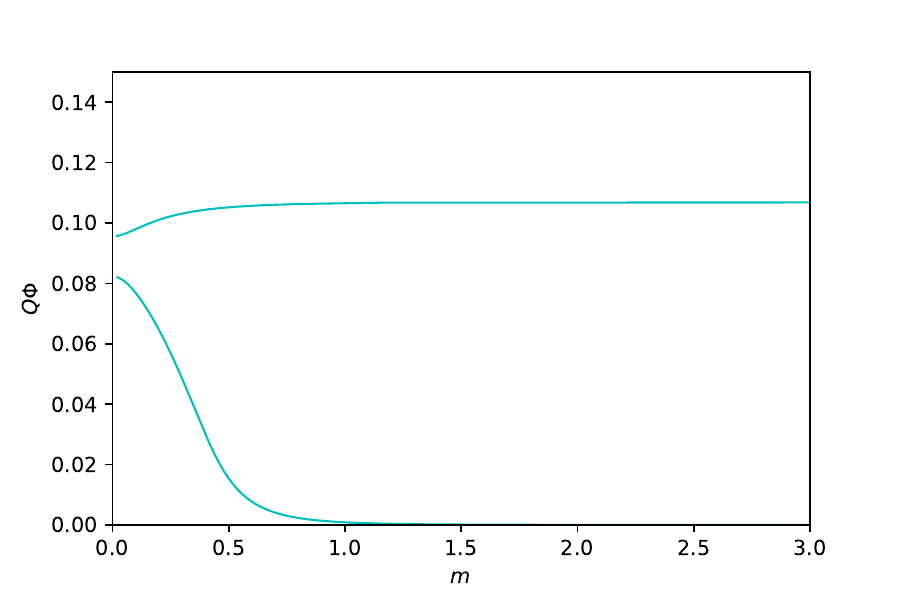}
\end{center}
\caption{The upper plot shows $\frac{\delta I}{\delta t}$  versus $m$ for black hole with single horizon. The parameters are fixed as $m_0=0.6, \Lambda=-1, \beta=0.2, d=6, k=1, c_1=c_2=c_3=c_4=-1.5$. The lower two plots shows $Q\Phi$ versus $m$ for black holes with inner and outer horizons. For a fixed $m$, the difference between the upper line and the lower line is action growth rate. The mass and electric charge parameters are fixed as $m_0=0.6, q=1.5$ (lower left plot) or $m_0=0.95, q=0.8$ (lower right plot), other parameters are fixed as $\Lambda=-1, \beta=0.2, d=6, k=1, c_1=c_2=c_3=c_4=-1.5$. }
\label{fig2}
\end{figure}

Note that, action growth  of the BI black holes in massive gravity (\ref{onemassive}) takes the identical form with that of EBI black holes\cite{1702.06766}.  However, graviton mass affects action growth through back-reaction on the geometry. The $m\rightarrow0$ limit of (\ref{massiveBH}) leads to black holes of EBI gravity
\begin{align}
\hat{f}(r)=k-\frac{m_0}{r^{d_3}}+\frac{4\beta^2-2\Lambda}{d_1d_2}r^2-\frac{4\beta^2r^2}{d_1d_2}\sqrt{1+\Gamma}+\frac{4d_2q^2}{d_1r^{2d_3}}\mathcal{H}.\label{EBIBH}
\end{align}
 When  mass and electric charge of the black hole are fixed, we have $f(r)-\hat{f}(r)<0$ (given $c_i$ negative\cite{1409.2369}) due to the graviton mass, which implies $r_{+}>\hat{r}_{+}$ ($\hat{r}_{+}$ is the horizon of the EBI black hole (\ref{EBIBH})),  therefore graviton mass leads BI black hole in massive gravity to complexitify faster than their Einstein gravity counterparts, just as the upper plot in Fig.\ref{fig2} shows.

It is easy to note that, when $q\rightarrow0$, action growth rate (\ref{onemassive}) reduces to the one of AdS-Schwarzschild black holes, for which Lloyd's bound is saturated (shown by the blue line of the upper plot in Fig.\ref{fig2}). Although the causal structure of single-horizoned BI black holes is identical with the one of AdS-Schwarzschild black holes, the manners of action growth  differ between the two types of black holes due to the presence of electromagnetic field. Therefore, BI electromagnetic field slows down complexification of the black holes.

For black holes with inner and outer horizons, the causal structure of which is shown by the right panel of Fig.\ref{fig1}. Bulk contributions to $\delta I$ are given by
\begin{align}
I_{\mathcal{V}_1}-I_{\mathcal{V}_2}=\frac{1}{16\pi}\Omega_{d_2}\delta t\left[\frac{4\Lambda}{d_1d_2}r^{d_1}-2F(r)-m^2\left(\frac{c_1}{d_2}r^{d_2}-c_3d_3r^{d_4}-2c_4d_3d_4r^{d_5}\right)\right]\bigg|_{r_{-}}^{r_{+}},\label{bulkm2}
\end{align}
The joint terms of action are
\begin{align}
I_{\mathcal{B}'\mathcal{B}}+I_{\mathcal{C}'\mathcal{C}}=&\frac{1}{16\pi}\Omega_{d_2}\delta t\bigg[-\frac{4\Lambda}{d_1d_2}r^{d_1}-d_3F(r)+rF'(r)\nonumber\\
&+m^2\left(\frac{c_1}{d_2}r^{d_2}-c_3d_3r^{d_4}-2c_4d_3d_4r^{d_5}\right)\bigg]\bigg|_{r_{-}}^{r_{+}}.\label{jointm2}
\end{align}
The total variation of action is then given by the sum of Eqs.(\ref{bulkm2}) and (\ref{jointm2}), which yields
\begin{align}
\frac{\delta I}{\delta t}=\left(M-Q\Phi\right)_{+}-\left(M-Q\Phi\right)_{-}.\label{twomassive}
\end{align}
Just as the single horizon case, this result is formally identical with the one of EBI black holes. Similarly, we also have $f(r)-\hat{f}(r)<0$ for fixed mass and charge parameters, which implies $r_{-}<\hat{r}_{-}$ and $r_{+}>\hat{r}_{+}$ ($\hat{r}_{\pm}$ are the inner and outer horizosn of the EBI black hole (\ref{EBIBH})), i.e., action growth rates of the double-horizoned BI black holes  in Einstein massive gravity are superior to the ones of the Einstein gravity counterparts too, just as the lower two plots in Fig.\ref{fig2} show us.

The lower left plot in Fig.\ref{fig2} shows the case that there exists an extremal black hole for the selected parameters, the left most point $A$ represents the extremal black hole for which the inner and outer horizons merge. The difference between the line above $A$ and the line under $A$ is action growth rate of the black hole with double horizons.  The lower right plot in Fig.\ref{fig2} shows the case that no extremal black hole exists for the selected parameters. It is easy to see that action growth rates increase as $m$ increases.

Now, we add a boundary term of electromagnetic field to the total action
\begin{align}
I_{\mu Q}&=\frac{\gamma}{16\pi}\int d\Sigma_\mu\frac{4F^{\mu\nu}}{\sqrt{1+\frac{F^2}{2\beta^2}}}A_\nu\nonumber\\
&=\frac{\gamma}{8\pi}\int d^dx\sqrt{-g}\frac{1}{\sqrt{1+\frac{F^2}{2\beta^2}}}F^{\mu\nu}F_{\mu\nu},
\end{align}
which does not alter the field equations. In the second equality we have used Stokes' theorem and the field equations $\nabla_\mu\big(\frac{F^{\mu\nu}}{\sqrt{1+\frac{F^2}{2\beta^2}}}\big)=0$. For BI black holes with single horizon, action growth becomes
\begin{align}
\frac{\delta I}{\delta t}=2M-(1-\gamma)Q\Phi-C_1,\label{onemassiveadd}
\end{align}
with
\begin{align}
C_1=\frac{\beta^2d_3^2}{8\pi d_1}\left(\frac{d_2}{d_3}\right)^{\frac{d_1}{2d_2}}\left(\frac{q}{\beta}\right)^{\frac{d_1}{d_2}}\left(d_3^{\frac{1}{d_2}}+\gamma\frac{d_1}{d_2}\right)\frac{\Gamma\left(1/(2d_2)\right)\Gamma\left(d_3/(2d_2)\right)}{\Gamma(1/2)}.
\end{align}
 For BI black holes with double horizons, action growth becomes
\begin{align}
\frac{\delta I}{\delta t}=\left[M-(1-\gamma)Q\Phi\right]_{+}-\left[M-(1-\gamma)Q\Phi\right]_{-}.\label{twomassiveadd}
\end{align}
Note that, if we set $\gamma=1$, $Q\Phi$ does not appear in the expressions of action growth (\ref{onemassiveadd}) and (\ref{twomassiveadd}). In this case, for BI black holes with single horizon, electric charge affects action growth only through  the constant $C_1$, for BI black holes with double horizons, action growth rates vanish. This agrees with the result obtained for dyonic black holes in Einstein-Maxwell gravity.

\begin{figure}[h]
\begin{center}
\includegraphics[width=.55\textwidth]{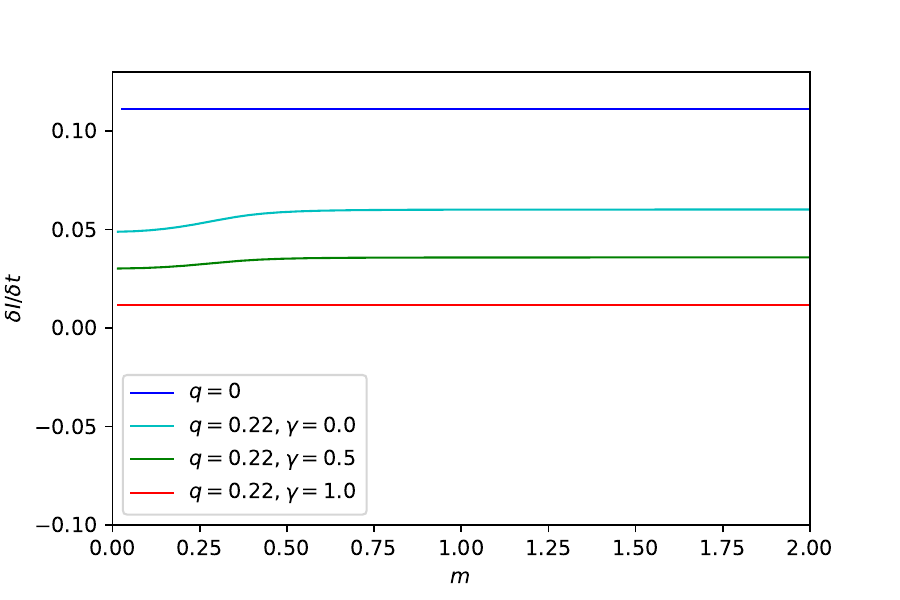}
\includegraphics[width=.45\textwidth]{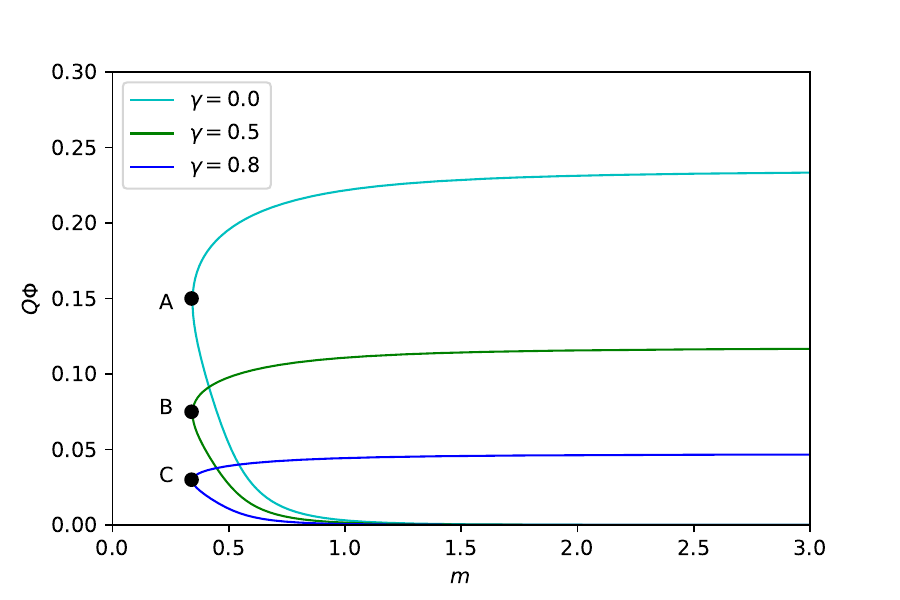}
\includegraphics[width=.45\textwidth]{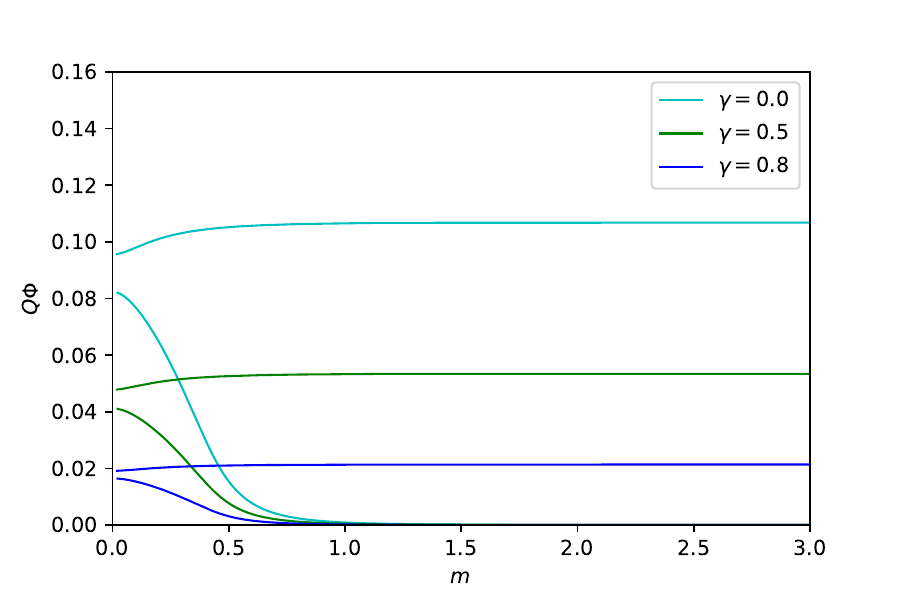}
\end{center}
\caption{The upper plot presents $\frac{\delta I}{\delta t}$ versus $m$ for the black holes with single horizon. The parameters are fixed as $m_0=0.7, \Lambda=-1, \beta=0.2, d=6, k=1, c_1=c_2=c_3=c_4=-1.5$. The lower two plots present $Q\Phi$ versus $m$ for black holes with inner and outer horizons. For a fixed $m$, the difference between the upper line and the lower line is action growth rate. The mass and electric charge parameters are fixed as $m_0=0.6, q=1.5$ (lower left plot) or $m_0=0.95, q=0.8$ (lower right plot), other parameters are fixed as $\Lambda=-1, \beta=0.2, d=6, k=1, c_1=c_2=c_3=c_4=-1.5$.}
\label{fig3}
\end{figure}

After addition of the BI boundary term, action growths versus $m$ are presented in Fig.\ref{fig3} for different values of $\gamma$. The straight blue line on the upper plot in Fig.\ref{fig3} still corresponds to the Lloyd's bound. The straight red line shows action growth of the black hole with single horizon for  $\gamma=1$. The difference between the straight blue line and the straight red line is just $C_1$. The three points $A, B, C$ on the lower left plot in Fig.\ref{fig3} still represent the extremal black holes. From the three plots in Fig.\ref{fig3}, one sees that, action growth rates decrease as $\gamma$ increases, and that action growth rates  increase as $m$ increases.

\section{EBI black holes\label{section4}}
The Lagrangian (\ref{LFm}) is only suitable for constructing BI black holes with electric charge. To construct black holes with both electric and magnetic charges, the Lagrangian (\ref{LFm}) should be replaced with the general one\cite{1606.02733}
\begin{align}
\mathcal{L}=-\beta^2\sqrt{-\det\left(g_{\mu\nu}+\frac{F_{\mu\nu}}{\beta}\right)}+\beta^2\sqrt{-\det(g_{\mu\nu})}.\label{BIlag}
\end{align}
In the large $\beta$ limit, (\ref{BIlag}) reduces to Maxwell theory. The Lagrangian of EBI theory is now given by
\begin{align}
\mathcal{L}=\sqrt{-g}(R-2\Lambda_0)-\beta^2\sqrt{-\det\left(g_{\mu\nu}+\frac{F_{\mu\nu}}{\beta}\right)},\label{EBI}
\end{align}
where $\Lambda_0=\Lambda-\beta^2/2$ is the bare cosmological constant, and $\Lambda$ is the effective cosmological constant. Variation of the action of EBI theory give rise to  field equations
\begin{align}
G^{\mu\nu}+&\Lambda_0g^{\mu\nu}+\frac{\beta^2}{2}\frac{\sqrt{-h}}{\sqrt{-g}}\left(h^{-1}\right)^{(\mu\nu)}=0,\label{eomdyong}\\
&\nabla_\mu\left[\frac{\sqrt{-h}}{\sqrt{-g}}\beta\left(h^{-1}\right)^{[\mu\nu]}\right]=0,\label{eomem}
\end{align}
where $h_{\mu\nu}=g_{\mu\nu}+F_{\mu\nu}/\beta$, $h\equiv\det(h_{\mu\nu})$, and $\left(h^{-1}\right)^{\mu\nu}$ denotes the inverse of $h_{\mu\nu}$, i.e.,
\begin{align}
\left(h^{-1}\right)^{\mu\rho}h_{\rho\nu}=\delta^\mu_\nu,\;\;\;\;\;\;\;\;h_{\nu\rho}\left(h^{-1}\right)^{\rho\mu}=\delta^\mu_\nu.
\end{align}

We still take the metric and strength ansatz given in Eq.(\ref{strengthansatz}). Solving e.o.m of $A_\mu$ (\ref{eomem}), we have
\begin{align}
\Phi'(r)=\frac{q}{\sqrt{\left(r^4+\frac{p^2}{\beta^2}\right)^n+\frac{q^2}{\beta^2}}},
\end{align}
where "$'$" denotes derivative with respect to $r$. The Einstein equations (\ref{eomdyong}) imply the dyonic  black hole solution in general dimensions\cite{1606.02733}
\begin{align}
f(r)=-\frac{\mu}{r^{2n-1}}-\frac{\Lambda_0}{n(2n+1)}r^2-\frac{\beta^2}{2n r^{2n-1}}G(r),\label{generalBH}
\end{align}
where $\mu$ is the mass parameter, and
\begin{align}
G(r)\equiv \int dr \sqrt{\left(r^4+\frac{p^2}{\beta^2}\right)^n+\frac{q^2}{\beta^2}}\label{G0}
\end{align}
 is introduced for convenience. The details of geometry and thermodynamics of the black hole (\ref{generalBH}) can be found in \cite{1606.02733}. The integral (\ref{G0}) can not be integrated out for general $n$, let's consider the following special cases.

\subsection{Purely electric EBI black holes\label{section41}}
If we take the $p\rightarrow 0$ limit of the dyonic black holes (\ref{generalBH}), we obtain the  purely electric EBI black holes.
For  electric EBI black holes with single horizon,  with e.o.m (\ref{eomdyong}) and the null coordinates introduced in (\ref{nullcoord}), at late times the bulk contributions to action growth are given by
\begin{align}
I_{\mathcal{V}_1}-I_{\mathcal{V}_2}&=\frac{1}{16\pi}\omega_2^n\delta t\left[\frac{2\Lambda_0}{n(2n+1)}r_{+}^{2n+1}+\frac{\beta^2}{n}\hat{G}(r_{+})-\frac{\beta^2}{n}\hat{G}(0)\right].\label{elecV1V2}
\end{align}
where $\hat{G}(r)\equiv\int dr \sqrt{r^{4n}+\frac{q^2}{\beta^2}}$
is $G(r)$ in (\ref{G0}) with the magnetic charge parameter $p=0$.
Action of the $r=0$ surface is
\begin{align}
I_{\mathcal{S}}
=\frac{1}{16\pi}\omega_2^n\delta t\left[(2n+1)\mu-q^{\frac{2n+1}{2n}}\beta^{\frac{2n-1}{2n}}\frac{\Gamma\left(1/2-1/(4n)\right)\Gamma\left(1+1/(4n)\right)}{\Gamma\left(1/2\right)}\right].\label{elecK}
\end{align}
Joint terms in the action are given by
\begin{align}
I_{\mathcal{B}'\mathcal{B}}
=\frac{\omega_2^n\delta t}{16\pi}\left[(2n-1)\mu-\frac{2\Lambda_0}{n(2n+1)}r_{+}^{2n+1}+\frac{2n-1}{2n}\beta^2\hat{G}(r_{+})-\frac{\beta^2r_{+}}{2n}\sqrt{r_{+}^{4n}+\frac{q^2}{\beta^2}}\right].\label{elecjoint}
\end{align}
By sum of Eqs. (\ref{elecV1V2}), (\ref{elecK}) and (\ref{elecjoint}) we have the action growth rate
\begin{align}
\frac{\delta I}{\delta t}&=\frac{\omega_2^n}{16\pi}\left[4n\mu+\frac{2n+1}{2n}\beta^2\hat{G}(r_{+})-\frac{\beta^2r_{+}}{2n}\sqrt{r_{+}^{4n}+\frac{q^2}{\beta^2}}-\frac{\beta^2}{n}\hat{G}(0)\right]\nonumber\\
&=2M-Q_e\Phi_e-\hat{C},\label{onepureelec}
\end{align}
with
\begin{align}
M=\frac{n\omega_2^n}{8\pi}\mu,\;\;\;\;\;Q_e=\frac{q}{16\pi}\omega_2^n,\;\;\;\;\;\Phi_e=\int_{r_{+}}^\infty\frac{qdr}{\sqrt{r^{4n}+\frac{q^2}{\beta^2}}}
\end{align}
are  mass, electric charge and  potential respectively, and the constant
\begin{align}
\hat{C}=\frac{(2n-1)\omega_2^n q^{\frac{2n+1}{2n}}\beta^{\frac{2n-1}{2n}}}{16\pi(2n+1)}\frac{\Gamma\left(1/2-1/(4n)\right)\Gamma\left(1+1/(4n)\right)}{\Gamma\left(1/2\right)}.\label{Celectric}
\end{align}
We see that, action growth of the electric EBI black holes is formally identical with that of the massive gravity counterparts.  BI electromagnetic field does not change casual structure of the spacetime compared to AdS-Schwarzschild black holes, but it affects the manner of action growth, which leads Lloyd's bound to be unsaturated.

\begin{figure}[h]
\begin{center}
\includegraphics[width=.45\textwidth]{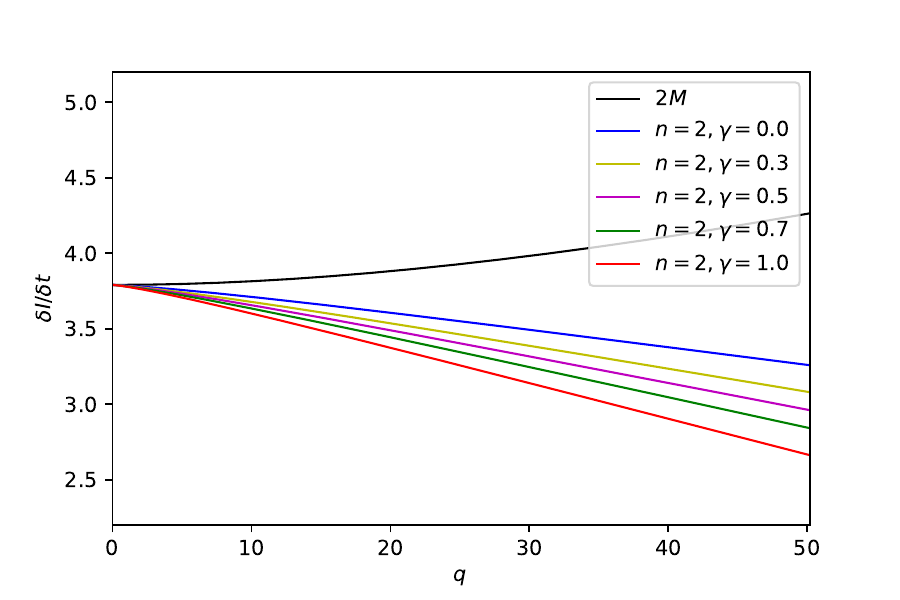}
\includegraphics[width=.45\textwidth]{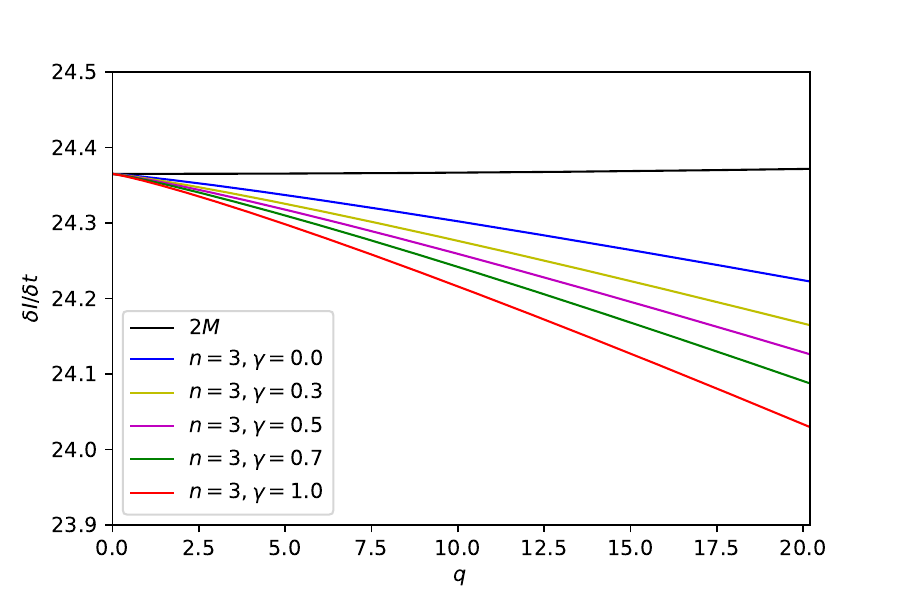}
\end{center}
\caption{$\frac{\delta I}{\delta t}$ versus $q$ for the purely electric black holes with single horizon. The black line on each plot is the $2M$ line. The parameters are fixed as $ \beta=0.2, r_{+}=3, \Lambda=-1$.}
\label{fig4}
\end{figure}

For electric EBI black holes  with double horizons,  the bulk contributions to $\delta I$ are
\begin{align}
I_{\mathcal{V}_1}-I_{\mathcal{V}_2}=\frac{1}{16\pi}\omega_2^n\delta t\left[\frac{2\Lambda_0}{n(2n+1)}r^{2n+1}+\frac{\beta^2}{n}\hat{G}(r)\right]\bigg|_{r_{-}}^{r_{+}}.\label{elec2bulk}
\end{align}
The joint terms of action are given by
\begin{align}
I_{\mathcal{B}'\mathcal{B}}+I_{\mathcal{C}'\mathcal{C}}=\frac{\omega_2^n\delta t}{16\pi}\left[(2n-1)\mu-\frac{2\Lambda_0}{n(2n+1)}r^{2n+1}+\frac{2n-1}{2n}\beta^2\hat{G}(r)-\frac{\beta^2r}{2n}\sqrt{r^{4n}+\frac{q^2}{\beta^2}}\right]\bigg|_{r_{-}}^{r_{+}}.
\label{elec2joint}
\end{align}
Combining (\ref{elec2bulk}) and (\ref{elec2joint}) we have the total action growth rate
\begin{align}
\frac{\delta I}{\delta t}=\left(M-Q_e\Phi_e\right)_{+}-\left(M-Q_e\Phi_e\right)_{-}.\label{twopureelec}
\end{align}
In this case, action growth of electric EBI black holes takes the identical form with that of the AdS-RN black holes, however, electromagnetic field affects action growth through the nonlinearity of BI theory.

We add a boundary term of electromagnetic field to the  action
\begin{align}
I_{\mu Q}&=\frac{\gamma}{16\pi}\int d\Sigma_\mu\beta\frac{\sqrt{-h}}{\sqrt{-g}}\left(h^{-1}\right)^{[\nu\mu]}A_\nu\nonumber\\
&=\frac{\gamma}{32\pi}\int d^dx\beta\sqrt{-h}\left(h^{-1}\right)^{[\nu\mu]}F_{\mu\nu},\label{boundaryBI}
\end{align}
which does not affect the field equations. The field equations $\nabla_\mu\big(\frac{\sqrt{-h}}{\sqrt{-g}}\left(h^{-1}\right)^{[\nu\mu]}\big)=0$ and Stoke's theorem have been used in the second equality. For electric EBI black holes with single horizon, action growth becomes
\begin{align}
\frac{\delta I}{\delta t}&=2M-(1-\gamma)Q_e\Phi_e-\hat{C}_1,\label{onepureelecadd}
\end{align}
with
\begin{align}
\hat{C}_1=\left(\frac{2n-1}{2n+1}+\gamma\right)\frac{\omega_2^n q^{\frac{2n+1}{2n}}\beta^{\frac{2n-1}{2n}}}{16\pi}\frac{\Gamma\left(1/2-1/(4n)\right)\Gamma\left(1+1/(4n)\right)}{\Gamma\left(1/2\right)}.\label{Celectricadd}
\end{align}

Fig.\ref{fig4} shows us  $\frac{\delta I}{\delta t}$ versus $q$. Since it is difficult to work out $q$ as $q(r_{+})$, and subsequently $\frac{\delta I}{\delta t}$ as $\frac{\delta I}{\delta t}(r_{+})$, from (\ref{generalBH}) for fixed $\mu$, we fix $r_{+}$, then $M$ and $\frac{\delta I}{\delta t}$ are functions of $q$. The black line on each plot in Fig.\ref{fig4} is the $2M$ line. From Fig.\ref{fig4}, one sees that action growth rates decrease as $q$ or $\gamma$ increases, Lloyd's bound is satisfied. Note that $\frac{\delta I}{\delta t}$ will not be zero as $q$ increases, since for certain value of the black hole mass $M$, $q$ can't be arbitrarily large to ensure the existence of the black hole horizon.

For electric EBI black holes with double horizons, action growth becomes
\begin{align}
\frac{\delta I}{\delta t}=\left[M-(1-\gamma)Q_e\Phi_e\right]_{+}-\left[M-(1-\gamma)Q_e\Phi_e\right]_{-}.\label{twopureelecadd}
\end{align}
 As we will see in the following,  action growth of the electric EBI black holes with $\gamma=1$ is in the same manner as that of the magnetic EBI black holes with $\gamma=0$, and vice versa.

\subsection{Pure magnetic EBI black holes\label{section42}}
In this subsection, we calculate action growth of purely magnetic EBI black hole in general dimensions and make a comparison between the effects of electric and magnetic charges on action growth.

For magnetic EBI black holes with single horizon, with e.o.m (\ref{eomdyong}) and the null coordinates, we obtain the bulk contributions to $\delta I$
\begin{align}
I_{\mathcal{V}_1}-I_{\mathcal{V}_2}&=\frac{1}{16\pi}\omega_2^n\delta t\left[\frac{2\Lambda_0}{n(2n+1)}r^{2n+1}+\frac{\beta^2}{n}\bar{G}(r)-p^2\bar{H}(r)\right]\bigg|_\epsilon^{r_{+}},\nonumber\\
&=\frac{1}{16\pi}\omega_2^n\delta t\left[\frac{2\Lambda_0}{n(2n+1)}r_{+}^{2n+1}+\frac{\beta^2}{n}\bar{G}(r_{+})-p^2\bar{H}(r_{+})-\frac{\beta^2}{n}\bar{G}(0)+p^2\bar{H}(0)\right].\label{magV1V2}
\end{align}
where $\bar{G}(r)\equiv\int dr\left(r^4+\frac{p^2}{\beta^2}\right)^{n/2},\;\bar{H}(r)\equiv \int dr\left(r^4+\frac{p^2}{\beta^2}\right)^{n/2-1}$, and we have taken  the $\epsilon\rightarrow0$ limit in the second equality.
Action of the spacelike surface is given by
\begin{align}
I_{\mathcal{S}}
=\frac{1}{16\pi}\omega_2^n\delta t\left[(2n+1)\mu+\frac{p^{n+1/2}}{2 n\beta^{n-3/2}}\frac{\Gamma\left(1/4\right)\Gamma\left(3/4-n/2\right)}{\Gamma\left(-n/2\right)}\right].\label{magK}
\end{align}
The joint terms in action are
\begin{align}
I_{\mathcal{B}'\mathcal{B}}
=\frac{\omega_2^n\delta t}{16\pi}\left[(2n-1)\mu-\frac{2\Lambda_0}{n(2n+1)}r_{+}^{2n+1}+\frac{2n-1}{2n}\beta^2\bar{G}(r_{+})-\frac{\beta^2r_{+}}{2n}\sqrt{(r_{+}^4+p^2/\beta^2)^n}\right].\label{magjoint}
\end{align}
We have the total action growth by sum of Eqs. (\ref{magV1V2}), (\ref{magK}) and (\ref{magjoint})
\begin{align}
\frac{\delta I}{\delta t}&=\frac{\omega_2^n}{16\pi}\left[4n\mu+\frac{2n+1}{2n}\beta^2\bar{G}(r_{+})-\frac{\beta^2r_{+}}{2n}\sqrt{(r_{+}^4+p^2/\beta^2)^n}-p^2\bar{H}(r_{+})-\frac{\beta^2}{n}\bar{G}(0)
+p^2\bar{H}(0)\right]\nonumber\\
&=2M-\bar{C},\label{onepuremag}
\end{align}
with
\begin{align}
M=\frac{n\omega_2^n}{8\pi}\mu,
\end{align}
and
\begin{align}
\bar{C}=-\frac{\omega_2^np^{n+1/2}}{8\pi(2n+1)\beta^{n-3/2}}\frac{\Gamma\left(1/4\right)\Gamma\left(3/4-n/2\right)}{\Gamma\left(-n/2\right)}.\label{Cmag}
\end{align}
Note that, if we don't consider the BI boundary term,  action growth of magnetic EBI black holes with single horizon depends only on mass and some constant involving $p$, it does not depend on the product of magnetic charge and potential $Q_m\Phi_m$.  This differs from the purely electric case (\ref{onepureelec}), where the action growth depends on mass, some constant  and the product $Q_e\Phi_e$.
By comparison of the first line of (\ref{onepuremag}) and the first line of (\ref{onepureelec}), we see that, for electric EBI black holes, the term corresponding to $\bar{H}(r_{+})$ disappears since $p$ vanishes, therefore the term $Q_e\Phi_e$ in action growth remains. The calculation details of simplifying Eq.(\ref{onepuremag}) from the first line to the second line can be found in the appendix.

It's interesting to note that, $\bar{C}$ is a positive constant for $n$ to be odd, while  $\bar{C}$ vanishes for  $n$ to be even. For odd $n$, according to CA duality we have $\frac{d\mathcal{C}}{dt}<\frac{2M}{\pi\hbar}$,  Lloyd's bound is satisfied. For even $n$, we have $\frac{d\mathcal{C}}{dt}=\frac{2M}{\pi\hbar}$, Lloyd's bound is saturated. Therefore, if we don't consider the boundary term of electromagnetic field, action growth of magnetic EBI black holes is in the same manner as the one of AdS-Schwarzschild black holes in some dimensions. In this case, magnetic charge affects action growth through back-reaction on the geometry.

\begin{figure}[h]
\begin{center}
\includegraphics[width=.45\textwidth]{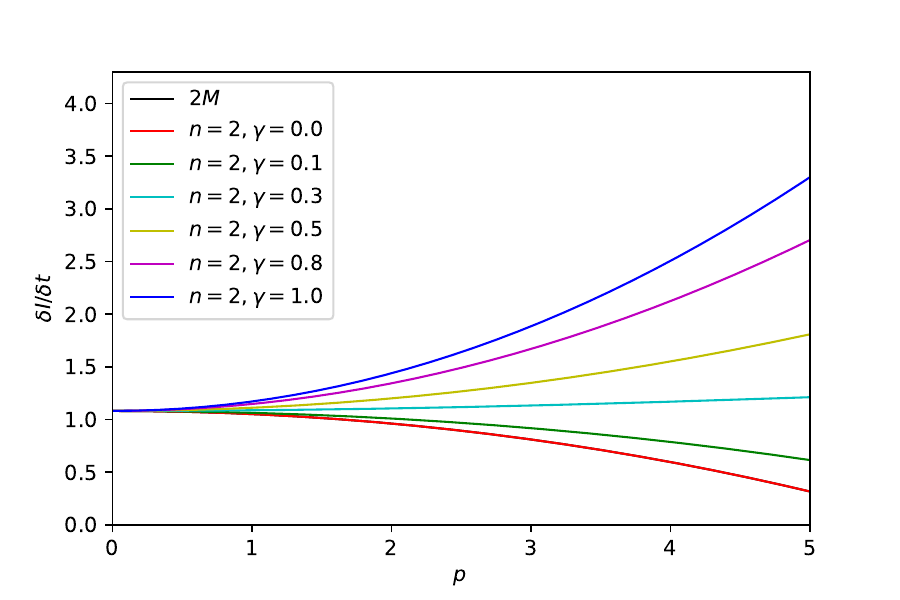}
\includegraphics[width=.45\textwidth]{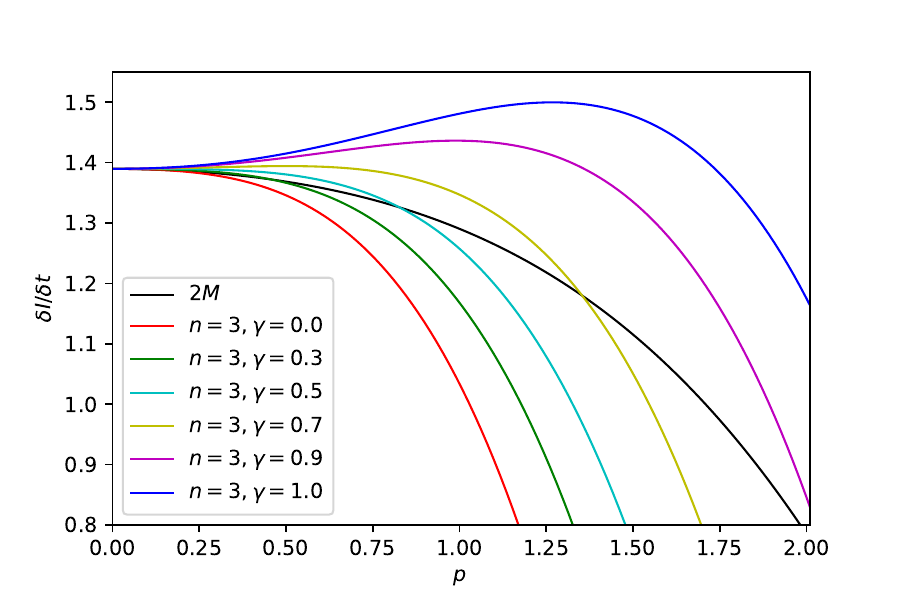}
\includegraphics[width=.45\textwidth]{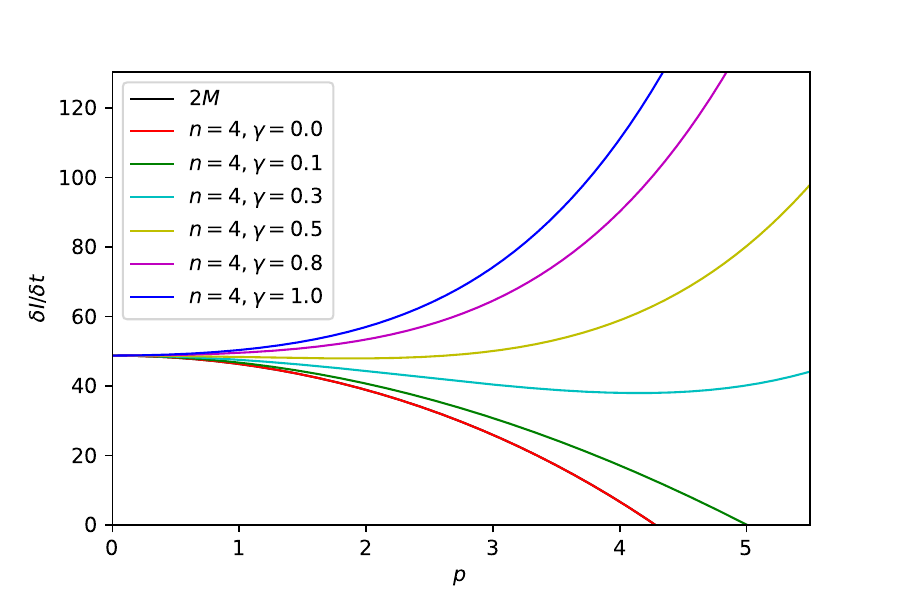}
\includegraphics[width=.45\textwidth]{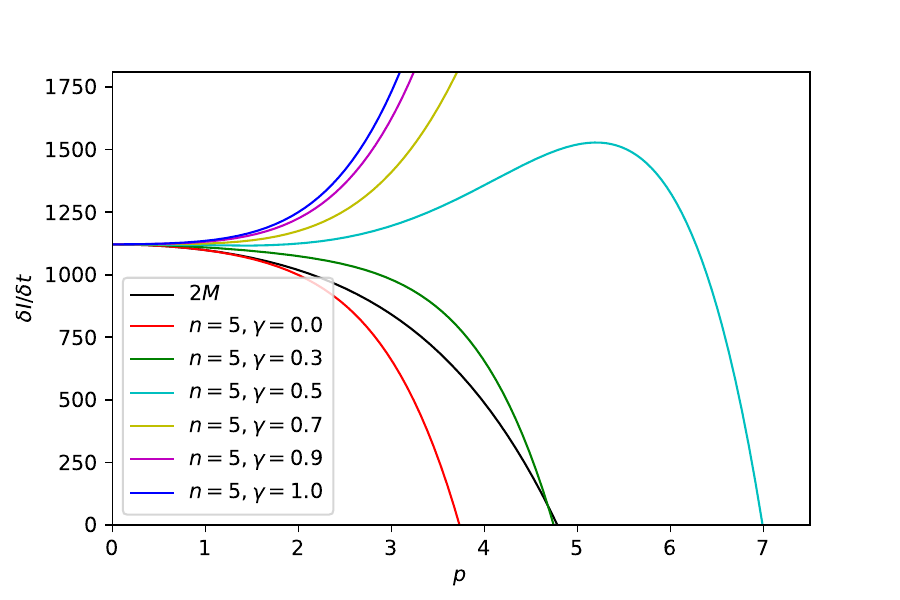}
\end{center}
\caption{$\frac{\delta I}{\delta t}$ versus $p$ for the purely magnetic black holes with single horizon. The black line on each plot is the $2M$ line. For the $n=2$ and $n=4$ cases, the black lines coincide with the red lines. The parameters are fixed as $ \beta=0.2, r_{+}=3, \Lambda=-1$ (upper left plot), $ \beta=0.3, r_{+}=2, \Lambda=-1$ (upper right plot), $ \beta=1.2, r_{+}=3, \Lambda=-1$ (lower left plot), $ \beta=0.5, r_{+}=3, \Lambda=-1$ (lower right plot).}
\label{fig5}
\end{figure}

For magnetic EBI black holes  with double horizons,  the bulk contributions to $\delta I$ are
\begin{align}
I_{\mathcal{V}_1}-I_{\mathcal{V}_2}=\frac{1}{16\pi}\omega_2^n\delta t\left[\frac{2\Lambda_0}{n(2n+1)}r^{2n+1}+\frac{\beta^2}{n}\bar{G}(r)-p^2\bar{H}(r)\right]\bigg|_{r_{-}}^{r_{+}}.\label{mag2bulk}
\end{align}
The joint terms in action are
\begin{align}
I_{\mathcal{B}'\mathcal{B}}+I_{\mathcal{C}'\mathcal{C}}=\frac{\omega_2^n\delta t}{16\pi}\left[(2n-1)\mu-\frac{2\Lambda_0}{n(2n+1)}r^{2n+1}+\frac{2n-1}{2n}\beta^2\bar{G}(r)-\frac{\beta^2r}{2n}\sqrt{(r^4+p^2/\beta^2)^n}\right]\bigg|_{r_{-}}^{r_{+}}.
\label{mag2joint}
\end{align}
Combining (\ref{mag2bulk}) and (\ref{mag2joint}) we find that
\begin{align}
\frac{\delta I}{\delta t}=0.\label{twopuremag}
\end{align}
This result agrees with that of the magnetic black holes in Einstein-Maxwell theory. Although action growth rates do not vanish for the magnetic EBI black holes with single horizon, action growth rates vanish for the magnetic black holes with double horizons both in Einstein-Maxwell gravity and in EBI gravity.

We now add the boundary term of electromagnetic field (\ref{boundaryBI}) to the  action. For magnetic EBI black holes with single horizon, action growth becomes
\begin{align}
\frac{\delta I}{\delta t}=2M-\gamma Q_m\Phi_m-\bar{C}_1,\label{onepuremagadd}
\end{align}
with $Q_m=\frac{np}{16\pi}\omega_2,\;\Phi_m=\frac{\partial M}{\partial Q_m}$ are magnetic charge and potential respectively, and
\begin{align}
\bar{C}_1=-\frac{\omega_2^np^{n+1/2}}{8\pi\beta^{n-3/2}}\left(\frac{1}{2n+1}-\frac{\gamma}{4}\right)\frac{\Gamma\left(1/4\right)\Gamma\left(3/4-n/2\right)}{\Gamma\left(-n/2\right)}.\label{barC1}
\end{align}
After the addition of the BI boundary term, we see that, magnetic charge affects action growth in a similar manner to electric charge. Here, $\bar{C}_1$ also vanishes for even $n$. Therefore, in the dimensions with even $n$, action growth of magnetic EBI black holes takes the specific form $\frac{\delta I}{\delta t}=2M-\gamma Q_m\Phi_m$.

Fig.\ref{fig5} shows us $\frac{\delta I}{\delta t}$ versus $p$. Similar to the electric case, we fix $r_{+}$, the black line on each plot in Fig.\ref{fig5} is the $2M$ line. On the left two plots, the black lines coincide with the red lines, this is because $\bar{C}_1$ in (\ref{barC1}) vanishes for even $n$. One sees from the figure that,  Lloyd's bound may be violated for some values of $\gamma$.  Therefore, the addition of matter field boundary term to action may lead to violation of Lloyd's bound. The late-time violations of Lloyd's bound have also been found in Einstein-dilaton system, in Einstein-Maxwell-Dilaton system\cite{1808.09917,1712.09826}, etc. Note also that, $\frac{\delta I}{\delta t}$ will not be zero as $p$ increases, since $p$ can't be arbitrarily large for certain $M$ to ensure the existence of black hole horizon.


For magnetic BI black holes with double horizons, action growth becomes
\begin{align}
\frac{\delta I}{\delta t}=\left[(M-\gamma Q_m\Phi_m)\right]_+-\left[(M-\gamma Q_m\Phi_m)\right]_-.\label{twopuremagadd}
\end{align}
Now, the rate of action growth does not vanish due to addition of the BI boundary term. If we set $\gamma=1$, action growth of magnetic EBI black holes is formally identical with that of the electric ones with $\gamma=0$. If we set $\gamma=\frac{1}{2}$, the manners of  action growth are similar for magnetic and electric EBI black holes, which implies electric and magnetic charges may contribute to action growth on equal footing for the dyonic EBI black hole. As we will see in the next subsection, this is indeed the case.

\subsection{Four-dimensional dyonic EBI black hole\label{section43}}
Since the integral (\ref{G0}) can not be integrated out for general $n$, for simplicity we only consider the $n=1$ case, i.e., the dyonic black hole in four dimensions.

For the dyonic EBI black hole with single horizon, bulk contributions to $\delta I$ are given by
\begin{align}
I_{\mathcal{V}_1}-I_{\mathcal{V}_2}=\frac{1}{16\pi}\omega_2\delta t\left[\frac{2\Lambda_0}{3}r_{+}^{3}+\beta^2\tilde{G}(r_{+})-p^2\tilde{H}(r_{+})-\beta^2\tilde{G}(0)+p^2\tilde{H}(0)\right].\label{dyonV1V2}
\end{align}
with
\begin{align}
\tilde{G}(r)\equiv\int dr\sqrt{r^4+\frac{p^2+q^2}{\beta^2}},\;\;\;\;\;\tilde{H}(r)\equiv \int dr\frac{1}{\sqrt{r^4+\frac{p^2+q^2}{\beta^2}}}.
\end{align}
Action of the $r=0$ spacelike surface  is
\begin{align}
I_{\mathcal{S}}=\frac{1}{16\pi}\omega_2\delta t\left[3\mu-\frac{(p^2+q^2)^{3/4}\sqrt{\beta}}{\sqrt{\pi}}\Gamma\left(\frac{1}{4}\right)\Gamma\left(\frac{5}{4}\right)\right].\label{dyonK}
\end{align}
The joint terms in action are
\begin{align}
I_{\mathcal{B}'\mathcal{B}}=\frac{1}{16\pi}\omega_2\delta t\left[\mu-\frac{2\Lambda_0}{3}r_{+}^{3}+\frac{1}{2}\beta^2\tilde{G}(r_{+})-\frac{\beta^2r_{+}}{2}\sqrt{r_{+}^4+(p^2+q^2)/\beta^2}\right].\label{dyonjoint}
\end{align}
Adding up all the contributions, the growth rate of action can be written as
\begin{align}
\frac{\delta I}{\delta t}=2M-Q_e\Phi_e-\tilde{C},\label{onedyon}
\end{align}
where
\begin{align}
\tilde{C}=\frac{(4p^2+q^2)\omega_2\sqrt{\beta}}{48\pi\left(q^2+p^2\right)^{1/4}}\frac{\Gamma\left(\frac{1}{4}\right)\Gamma\left(\frac{5}{4}\right)}{\Gamma\left(\frac{1}{2}\right)},
\end{align}
and
\begin{align}
M&=\frac{\omega_2}{8\pi}\mu,\;\;\;\;\;\;\;\;\;\;\;\;\;\;Q_e=\frac{q}{16\pi}\omega_2,\nonumber\\
\Phi_e&=\frac{q}{r_{+}}
\sideset{_2}{_1}{\mathop{F}}\left[\frac{1}{4},\frac{1}{2},\frac{5}{4},-\frac{p^2+q^2}{r_{+}^4\beta^2}\right].
\end{align}
are mass, electric charge and potential  respectively.

The result (\ref{onedyon}) matches that of purely electric and magnetic EBI black holes with single horizon in general dimensions above, i.e., only the product of electric charge and potential appears in the expression of action growth, magnetic charge affects action growth only through some constant.

\begin{figure}[h]
\begin{center}
\includegraphics[width=.45\textwidth]{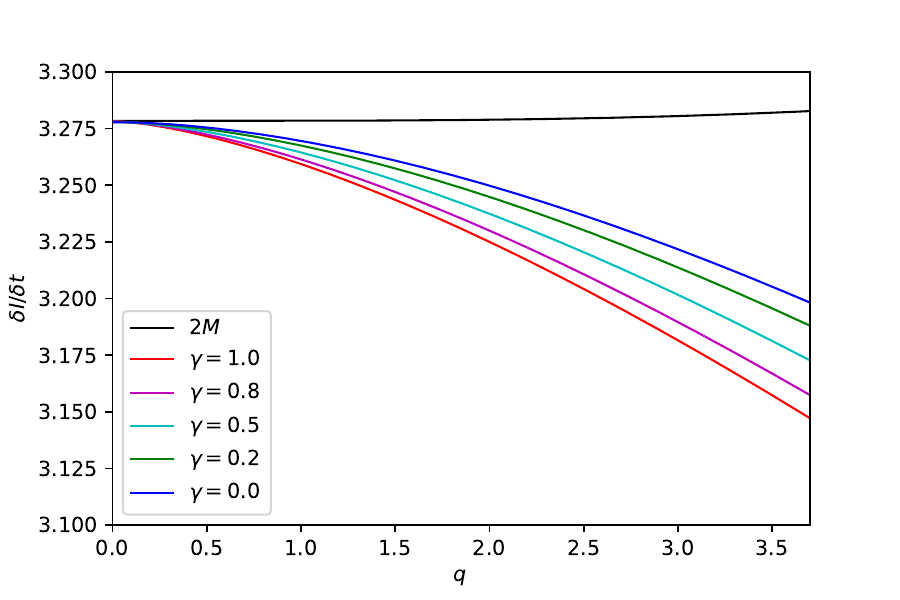}
\includegraphics[width=.45\textwidth]{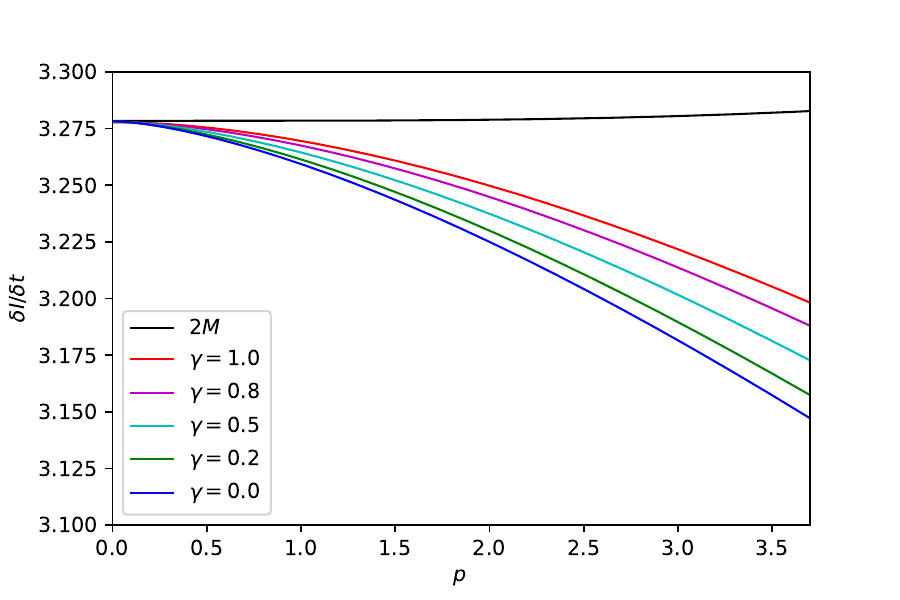}
\end{center}
\caption{$\frac{\delta I}{\delta t}$ versus $q$ with fixed $p$ (left plot) and $\frac{\delta I}{\delta t}$ versus $p$ with fixed $q$ (right plot) for four-dimensional dyonic black hole with single horizon. The parameters are fixed as $\beta=0.15, r_{+}=5, \Lambda=-1, p=0.1$ for the left plot, and $\beta=0.15, r_{+}=5, \Lambda=-1, q=0.1$ for the right plot.}
\label{fig6}
\end{figure}

For the dyonic EBI black hole with inner and outer horizons,  we have
\begin{align}
I_{\mathcal{V}_1}-I_{\mathcal{V}_2}&=\frac{1}{16\pi}\omega_2\delta t\left[\frac{2\Lambda_0}{3}r^{3}+\beta^2\tilde{G}(r)-p^2\tilde{H}(r)\right]\bigg|_{r_{-}}^{r_{+}},\\
I_{\mathcal{B}'\mathcal{B}}+I_{\mathcal{C}'\mathcal{C}}&=\frac{1}{16\pi}\omega_2\delta t\left[\mu-\frac{2\Lambda_0}{3}r^{3}+\frac{1}{2}\beta^2\tilde{G}(r)-\frac{\beta^2r}{2}\sqrt{r^4+(p^2+q^2)/\beta^2}\right]\bigg|_{r_{-}}^{r_{+}}.
\end{align}
Thus the growth rate of action is
\begin{align}
\frac{\delta I}{\delta t}&=\left(M-Q_e\Phi_e\right)_{+}-\left(M-Q_e\Phi_e\right)_{-}.\label{twodyon}
\end{align}
Similarly, only $Q_e\Phi_e$ appears in $\frac{\delta I}{\delta t}$. This result is formally identical with that of dyonic black holes in Einstein-Maxwell gravity. It seems that, independent of the linearity or  nonlinearity of electromagnetic theory, only electric charge affects action growth of the  black holes with double horizons, action growth rates vanish for the purely magnetically charged black holes.

After addition of  the boundary term (\ref{boundaryBI}), for dyonic EBI black holes with single horizon, action growth becomes
\begin{align}
\frac{\delta I}{\delta t}=2M-(1-\gamma)Q_e\Phi_e-\gamma Q_m\Phi_m-\tilde{C}_1,\label{onedyonadd}
\end{align}
where $Q_m$, $\Phi_m$ are magnetic charge and potential
\begin{align}
Q_m=\frac{p}{16\pi}\omega_2,\;\;\;\Phi_m=\frac{p}{r_{+}}
\sideset{_2}{_1}{\mathop{F}}\left[\frac{1}{4},\frac{1}{2},\frac{5}{4},-\frac{p^2+q^2}{r_{+}^4\beta^2}\right],
\end{align}
and
\begin{align}
\tilde{C}_1=\frac{\omega_2\sqrt{\beta}}{16\pi^{3/2}\left(q^2+p^2\right)^{1/4}}\left[\frac{5}{6}(p^2+q^2)+\left(\frac{1}{2}-\gamma\right)(p^2-q^2)\right]\frac{\Gamma\left(\frac{1}{4}\right)\Gamma\left(\frac{5}{4}\right)}
{\Gamma\left(\frac{1}{2}\right)}.\label{tildeC1}
\end{align}

Fig.\ref{fig6} shows $\frac{\delta I}{\delta t}$ versus $q$ and $\frac{\delta I}{\delta t}$ versus $p$. It is easy to see that, $\frac{\delta I}{\delta t}$ decreases with  $q$ (or $p$) for fixed $p$ (or $q$), and Lloyd's bound is satisfied. Note that, on the left plot in Fig.\ref{fig6}, the lines with larger $\gamma$ are lower than the lines with smaller $\gamma$, while, on the right plot in Fig.\ref{fig6}, the lines with larger $\gamma$ are upper than the lines with smaller $\gamma$. This is because the coefficient before $Q_e\Phi_e$ in Eq.(\ref{onedyonadd}) is $(1-\gamma)$  while the coefficient before $Q_m\Phi_m$ in Eq.(\ref{onedyonadd}) is $\gamma$.

For dyonic BI black holes with double horizons, action growth becomes
\begin{align}
\frac{\delta I}{\delta t}&=\left[M-(1-\gamma)Q_e\Phi_e-\gamma Q_m\Phi_m\right]_{+}-\left[M-(1-\gamma)Q_e\Phi_e-\gamma Q_m\Phi_m\right]_{-}.\label{twodyonadd}
\end{align}
Note from (\ref{onedyonadd}) and (\ref{twodyonadd}) that, if we set $\gamma=1$,  $Q_e\Phi_e$ in the expressions of action growth vanishes, only  $Q_m\Phi_m$ remains, electric charge  affects action growth of the dyonic EBI black hole with single horizon only through some constant. If we set $\gamma=\frac{1}{2}$, the asymmetric part under $p\leftrightarrow q$ in (\ref{tildeC1}) vanishes, then electric and magnetic charges contribute to action growth on equal footing for the four-dimensional dyonic EBI black holes.

\section{Summary and Discussion}
In this paper, we study action growth of BI black holes. As a comparison, we first review action growth of dyonic black holes in Einstein-Maxwell gravity in general dimensions, and notice that similar to the four-dimensional case,  action growth rates vanish for purely magnetic black holes if we don't consider the Maxwell boundary term. After the inclusion of Maxwell boundary term, if we set $\gamma=\frac{1}{2}$, then electric and magnetic charges contribute to action growth on equal footing, which is in accord with electric-magnetic duality. If we set $\gamma=1$, then action growth rates vanish for purely electric black holes.

We study action growth of electric BI black holes in massive gravity, and find that BI black holes in massive gravity always complexify faster than their Einstein gravity counterparts due to the back-reaction of graviton mass on geometry. If the BI boundary term is included, we find that action growth rates decrease as $\gamma$ increases.

For the EBI black holes, we first study action growths of the purely electric and magnetic black holes in general dimensions. Before considering the BI boundary term,  action growths of electric and magnetic EBI black holes with single horizon are
in different manners,
which are $\frac{\delta I}{\delta t}=2M-Q_e\Phi_e-\hat{C}$ and $\frac{\delta I}{\delta t}=2M-\bar{C}$ respectively.
We notice that, for the magnetic black holes, the constant $\bar{C}$ vanishes for even $n$. Therefore, in the dimensions with even $n$, action growth of magnetic EBI black holes saturates Lloyd's bound, i.e., it takes the identical form with that of the AdS-Schwarzschild black holes. In this case, magnetic charge affects action growth through back-reaction on the geometry. For EBI black holes with double hirizons, action growth of the electric ones takes the identical form with that of AdS-RN black holes, while action growth vanishes for the magnetic ones, this agrees with the result obtained from the dyonic black holes in Einstein-Maxwell gravity.

If we include the BI boundary term to the action, action growth of the electric (magnetic) EBI black holes with $\gamma=1$ takes  similar form with that of the magnetic (electric) ones with $\gamma=0$.
We find that,  for electric EBI black holes with single horizon Lloyd's bound is satisfied, while, for the magnetic ones Lloyd's bound may be violated for some values of $\gamma$. Therefore, the inclusion of BI boundary term to action may lead to the violation of Lloyd's bound.

For the four-dimensional dyonic EBI black hole, calculation shows the manner of action growth of the black hole agrees with that of the purely electric and magnetic EBI black holes in general dimensions. When we set $\gamma=\frac{1}{2}$, electric and magnetic charges contribute to action growth on equal footing. Lloyd's bound is satisfied for the four-dimensional dyonic EBI black hole.

\section*{Acknowledgment}

KM would like to thank Profs. Peng Wang, Haitang Yang, Liu Zhao and Haishan Liu for valuable discussions.

\section*{Appendix}

The magnetic potential is given by
\begin{align}
\Phi_m&=\frac{\partial M}{\partial Q_m}
=\frac{r_{+}^{2n+1}\beta^2\omega_2^{n-1}}{2np}\left(\left(1+\frac{p^2}{r_{+}^4\beta^2}\right)^{n/2}-
\sideset{_2}{_1}{\mathop{F}}\left[-\frac{1}{4}-\frac{n}{2},-\frac{n}{2},\frac{3}{4}-\frac{n}{2},-\frac{p^2}{r_{+}^4\beta^2}\right]\right).
\end{align}
The two terms in eq.(\ref{onepuremag}) can be simplified to
\begin{align}
\frac{\omega_2^n}{16\pi}\left(\frac{2n+1}{2n}\beta^2\bar{G}(r_{+})-\frac{\beta^2r_{+}}{2n}\sqrt{(r_{+}^4+p^2/\beta^2)^n}\right)=-\frac{1}{n}Q_m\Phi_m.\label{appendix2term}
\end{align}
From the definition of $\bar{H}(r_{+})$, we have
\begin{align}
\bar{H}(r_{+})&\equiv \int dr_{+}\left(r_{+}^4+\frac{p^2}{\beta^2}\right)^{n/2-1}
=\frac{r_{+}^{2n-3}}{2n-3}\sideset{_2}{_1}{\mathop{F}}\left[\frac{3}{4}-\frac{n}{2},1-\frac{n}{2},\frac{7}{4}-\frac{n}{2},-\frac{p^2}{r_{+}^4\beta^2}\right].
\end{align}
With the formula $\sideset{_2}{_1}{\mathop{F}}\left[a,b,c,z\right]=\frac{c-1}{(a-1)(b-1)}\frac{d}{dz}\sideset{_2}{_1}{\mathop{F}}\left[a-1,b-1,c-1,z\right]$, $\bar{H}(r_{+})$ can be rewritten as
\begin{align}
\bar{H}(r_{+})
=-\frac{r_{+}^{2n+1}\beta^2}{2np^2}\left(\left(1+\frac{p^2}{r_{+}^4\beta^2}\right)^{n/2}-
\sideset{_2}{_1}{\mathop{F}}\left[-\frac{1}{4}-\frac{n}{2},-\frac{n}{2},\frac{3}{4}-\frac{n}{2},-\frac{p^2}{r_{+}^4\beta^2}\right]\right).
\end{align}
Therefore
\begin{align}
\frac{\omega_2^n}{16\pi}p^2\bar{H}(r_{+})=-\frac{1}{n}Q_m\Phi_m.\label{appendixH}
\end{align}
By comparison of eqs.(\ref{appendix2term}) and (\ref{appendixH}), we immediately have
\begin{align}
\frac{\omega_2^n}{16\pi}\left(\frac{2n+1}{2n}\beta^2\bar{G}(r_{+})-\frac{\beta^2r_{+}}{2n}\sqrt{(r_{+}^4+p^2/\beta^2)^n}-p^2\bar{H}(r_{+})\right)=0.
\end{align}

\providecommand{\href}[2]{#2}\begingroup
\footnotesize\itemsep=0pt
\providecommand{\eprint}[2][]{\href{http://arxiv.org/abs/#2}{arXiv:#2}}


\begin{thebibliography}{}

\bibitem{0501128}
J. Erlich, E. Katz, D. T. Son, M. A. Stephanov,
``QCD and a holographic model of hadrons,''
Phys. Rev. Lett. $\mathbf{95}$, 261602 (2005) [\eprint{hep-ph/0501128}].

\bibitem{1101.2451}
I. Bredberg, C. Keeler, V. Lysov and A. Strominger,
``From Navier-Stokes To Einstein,''
JHEP $\mathbf{1207}$, 146 (2012) [\eprint{1101.2451}].

\bibitem{1103.3022}
G. Compere, P. McFadden, K. Skenderis, M. Taylor,
``The Holographic fluid dual to vacuum Einstein gravity,''
JHEP $\mathbf{1107}$, 050 (2011) [\eprint{1103.3022}].

\bibitem{0803.3295}
S. A. Hartnoll, C. P. Herzog, G. T. Horowitz,
``Building a Holographic Superconductor,''
Phys. Rev. Lett. $\mathbf{101}$, 031601 (2008) [\eprint{0803.3295}].

\bibitem{0810.1563}
S. A. Hartnoll, C. P. Herzog, G. T. Horowitz,
``Holographic Superconductors,''
JHEP $\mathbf{0812}$, 015 (2008) [\eprint{0810.1563}].

\bibitem{1306.0533}
J. Maldacena, L. Susskind,
``Cool horizons for entangled black holes,''
Fortsch. Phys. $\mathbf{61}$, 781 (2013) [\eprint{1306.0533}].

\bibitem{CV}
D. Stanford, L. Susskind,
``Complexity and Shock Wave Geometries,''
Phys. Rev. D $\mathbf{90}$, 126007 (2014) [\eprint{1406.2678}].


\bibitem{CV2}
L. Susskind, Y. Zhao,
``Switchbacks and the Bridge to Nowhere,''
 [\eprint{1408.2823}].

\bibitem{1711.10887}
S. Karar, S. Gangopadhyay,
``Holographic complexity for Lifshitz system,''
Phys. Rev. D $\mathbf{98}$, 026029 (2018) [\eprint{1711.10887}].

\bibitem{1803.06680}
B. Chen, W. M. Li, R. Q. Yang, C. Y. Zhang, S. J. Zhang,
``Holographic subregion complexity under a thermal quench,''
JHEP $\mathbf{1807}$, 034 (2018) [\eprint{1803.06680}].

\bibitem{1807.06361}
A. Bhattacharya, S. Roy,
``Holographic Entanglement Entropy, Subregion Complexity and Fisher Information metric of 'black' Non-SUSY D3 Brane,''
 [\eprint{1807.06361}].

\bibitem{1803.08627}
L. P. Du, S. F. Wu, H. B. Zeng,
``Holographic complexity of the disk subregion in (2+1)-dimensional gapped systems,''
Phys. Rev. D $\mathbf{98}$, 066005 (2018) [\eprint{1803.08627}].

\bibitem{1808.08719}
S. J. Zhang,
``Subregion complexity and confinement-deconfinement transition in a holographic QCD model,''
 [\eprint{1808.08719}].


\bibitem{1808.10169}
Y. Ling, Y. Liu,  C. Y. Zhang
``Holographic Subregion Complexity in Einstein-Born-Infeld theory,''
 [\eprint{1808.10169}].

\bibitem{1509.07876}
A. R. Brown, D. A. Roberts, L. Susskind, B. Swingle, Y. Zhao,
``Complexity Equals Action,''
Phys. Rev. Lett. $\mathbf{116}$, 191301 (2015) [\eprint{1509.07876}].

\bibitem{1512.04993}
A. R. Brown, D. A. Roberts, L. Susskind, B. Swingle, Y. Zhao,
``Complexity, action, and black holes,''
Phys. Rev. D $\mathbf{93}$, 086006 (2016) [\eprint{1512.04993}].

\bibitem{9310026}
G. 't Hooft,
``Dimensional reduction in quantum gravity,''
Conf. Proc. C $\mathbf{930308}$, 284 (1993) [\eprint{gr-qc/9310026}].

\bibitem{9409089}
L. Susskind,
``The World as a Hologram,''
J. Math. Phys. $\mathbf{36}$, 6377 (1995) [\eprint{hep-th/9409089}].

\bibitem{Bekenstein}
J. D. Bekenstein,
``A Universal Upper Bound on the Entropy to Energy Ratio for
Bounded Systems,''
Phys. Rev. D $\mathbf{23}$, 287 (1981).

\bibitem{9710043}
N. Margolus and L. B. Levitin,
``The Maximum speed of dynamical evolution,''
Physica D $\mathbf{120}$, 188 (1998) [\eprint{quant-ph/9710043}].

\bibitem{Lloyd}
S. Lloyd,
``Ultimate physical limits to computation,''
Nature $\mathbf{406}$, no. 6799, 1047 (2000).

\bibitem{1606.08307}
R. G.  Cai, S. M. Ruan, S. J. Wang, R. Q. Yang, R. H. Peng,
``Action growth for AdS black holes,''
JHEP $\mathbf{1609}$, 161 (2016) [\eprint{1606.08307}].

\bibitem{1702.06766}
R. G. Cai, M. Sasaki, S. J. Wang,
``Action growth of charged black holes with a single horizon,''
Phys. Rev. D $\mathbf{95}$, 124002 (2017) [\eprint{1702.06766}].

\bibitem{1703.06297}
J. Tao, P. Wang, H.  Yang,
``Testing holographic conjectures of complexity with Born-Infeld black holes,''
 Eur. Phys. J. C $\mathbf{77}$, no.12, 817 (2017) [\eprint{1703.06297}].

\bibitem{1703.10006}
P. Wang, H. Yang, S. Ying,
``Action growth in f(R) gravity,''
Phys. Rev. D $\mathbf{96}$, no.4, 046007 (2017) [\eprint{1703.10006}].

\bibitem{1712.09826}
B. Swingle, Y. Wang,
``Holographic Complexity of Einstein-Maxwell-Dilaton Gravity,''
JHEP $\mathbf{1809}$, 106 (2018) [\eprint{1712.09826}].

\bibitem{1703.10468}
W. D. Guo, S. W. Wei, Y. Y. Li, Y. X. Liu,
``Complexity growth rates for AdS black holes in massive gravity and f(R) gravity,''
Eur. Phys. J. C $\mathbf{77}$, no.12, 904 (2017) [\eprint{1703.10468}].


\bibitem{1612.03627}
W. J. Pan, Y. C. Huang,
``Holographic complexity and action growth in massive gravities,''
Phys. Rev. D $\mathbf{95}$, 126013 (2017) [\eprint{1612.03627}].

\bibitem{1801.03638}
Y. S. An, R. H. Peng,
``The effect of Dilaton on the holographic complexity growth,''
Phys. Rev. D $\mathbf{97}$, 066022 (2018) [\eprint{1801.03638}].

\bibitem{1806.06216}
R. Auzzi, S. Baiguera, M. Grassi, G. Nardelli, N. Zenoni,
``Complexity and action for warped AdS black holes,''
JHEP $\mathbf{1809}$, 013 (2018) [\eprint{1806.06216}].

\bibitem{1702.06796}
M. Alishahiha, A. F. Astaneh, A. Naseh, M. H. Vahidinia,
``On Complexity for Higher Derivative Gravities,''
JHEP $\mathbf{1705}$, 009 (2017) [\eprint{1702.06796}].

\bibitem{1806.10312}
J. Jiang, H. Zhang,
``Surface term, corner term, and action growth in F(Riemann) gravity theory,''
 [\eprint{1806.10312}].

 \bibitem{1706.03788}
A. Reynolds, S. F. Ross,
``Complexity in de Sitter Space,''
Class. Quant. Grav. $\mathbf{34}$, no.17, 175013 (2017) [\eprint{1706.03788}].


\bibitem{1803.02795}
P. A. Cano, R. A. Hennigar, H. Marrochio,
``Complexity Growth Rate in Lovelock Gravity,''
 Phys. Rev. Lett. $\mathbf{121}$, 121602 (2018) [\eprint{1803.02795}].


 \bibitem{1710.05686}
L. Sebastiani, L. Vanzo, S. Zerbini,
``Action growth for black holes in modified gravity,''
Phys. Rev. D $\mathbf{97}$, no.4, 044009 (2018)  [\eprint{1710.05686}].

\bibitem{1808.09917}
S. Mahapatra, P. Roy,
``On the time dependence of holographic complexity in a dynamical Einstein-dilaton model,''
  [\eprint{1808.09917}].

\bibitem{1610.05090}
R. Q.  Yang,
``Strong energy condition and complexity growth bound in holography,''
Phys. Rev. D $\mathbf{95}$, no.8, 086017 (2017)  [\eprint{1610.05090}].


\bibitem{1804.07410}
S. Chapman, H. Marrochio, R. C. Myers,
``Holographic Complexity in Vaidya Spacetimes I,''
JHEP $\mathbf{1806}$, 046 (2018)  [\eprint{1804.07410}].

\bibitem{1805.07262}
S. Chapman, H. Marrochio, R. C. Myers,
``Holographic Complexity in Vaidya Spacetimes II,''
JHEP $\mathbf{1806}$, 114 (2018)  [\eprint{1805.07262}].

 \bibitem{1810.00758}
J. Jiang,
``Action growth rate for a higher curvature gravitational theory,''
  [\eprint{1810.00758}].

\bibitem{1708.01779}
Y. G. Miao, L. Zhao,
``Complexity-action duality of the shock wave geometry in a massive gravity theory,''
Phys. Rev. D $\mathbf{97}$, no.2, 024035 (2018)  [\eprint{1708.01779}]..

\bibitem{1808.00067}
S. A. H. Mansoori,  V. Jahnke,  M. M. Qaemmaqami,  Y. D. Olivas
``Action growth rate for a higher curvature gravitational theory,''
  [\eprint{1808.00067}].


\bibitem{1311.7299}
D. C. Zou, S. J. Zhang, B. Wang,
``Critical behavior of Born-Infeld AdS black holes in the extended phase space thermodynamics,''
Phys. Rev. D $\mathbf{89}$, no.4, 044002 (2014) [\eprint{1311.7299}].

\bibitem{1712.08798}
K. Meng, D. B. Yang,
``Black holes of dimensionally continued gravity coupled to Born-Infeld electromagnetic field,''
Phys. Lett. B $\mathbf{780}$, 363 (2018) [\eprint{1712.08798}].

\bibitem{1804.10951}
K. Meng,
``Hairy black holes of Lovelock-Born-Infeld-scalar gravity,''
Phys. Lett. B $\mathbf{784}$, 56 (2018) [\eprint{1804.10951}].

\bibitem{1901.00014}
K. Goto, H. Marrochio, R. C. Myers, L. Queimada, B. Yoshida,
``Holographic Complexity Equals Which Action?,''
JHEP  $\mathbf{1902}$, 160 (2019) [\eprint{1901.00014}].

\bibitem{1905.06409}
H. S. Liu, H. Lu,
``Action Growth of Dyonic Black Holes and Electromagnetic Duality,''
[\eprint{1905.06409}].

\bibitem{1905.07576}
J. Jiang, M. Zhang,
``Holographic complexity of the electromagnetic black hole,''
[\eprint{1905.07576}].





\bibitem{1508.01311}
S. H. Hendi, B. E. Panah, S. Panahiyan,
``Einstein-Born-Infeld-Massive Gravity: adS-Black Hole Solutions and their Thermodynamical properties,''
JHEP $\mathbf{1511}$, 157 (2015) [\eprint{1508.01311}].

\bibitem{1609.00207}
L. Lehner, R. C. Myers, E. Poisson, R. D. Sorkin,
``Gravitational action with null boundaries,''
Phys. Rev. D $\mathbf{94}$, 084046 (2016) [\eprint{1609.00207}].


\bibitem{9403018}
D. Brill, G. Hayward,
``Is the gravitational action additive?'',
Phys. Rev. D $\mathbf{50}$, 4914 (1994) [\eprint{gr-qc/9403018}].

\bibitem{1409.2369}
R. G.  Cai, Y. P.  Hu, Q. Y. Pan, Y. L. Zhang,
``Thermodynamics of Black Holes in Massive Gravity,''
Phys.Rev. D $\mathbf{91}$, no.2, 024032 (2015) [\eprint{1409.2369}].

\bibitem{1606.02733}
S. Li, H. Lu, H. Wei,
``Dyonic (A)dS Black Holes in Einstein-Born-Infeld Theory in Diverse Dimensions,''
JHEP $\mathbf{1607}$, 004 (2016) [\eprint{1606.02733}].





\end{thebibliography}
\end{document}